\documentclass[prd,aps,showpacs,superscriptaddress,nofootinbib,preprint]{revtex4-1}
\usepackage{amssymb}
\usepackage{graphicx}
\def\slashchar#1{\setbox0=\hbox{$#1$}
   \dimen0=\wd0 \setbox1=\hbox{/} \dimen1=\wd1
   \ifdim\dimen0>\dimen1 \rlap{\hbox to \dimen0{\hfil/\hfil}} #1
   \else  \rlap{\hbox to \dimen1{\hfil$#1$\hfil}} / \fi}
\def\slashchar#1{\setbox0=\hbox{$#1$}
   \dimen0=\wd0 \setbox1=\hbox{/} \dimen1=\wd1
   \ifdim\dimen0>\dimen1 \rlap{\hbox to \dimen0{\hfil/\hfil}} #1
   \else  \rlap{\hbox to \dimen1{\hfil$#1$\hfil}} / \fi}

\begin{document}
\title{Effects of nuclear medium and nonisoscalarity in extracting $sin^2\theta_W$ using Paschos-Wolfenstein relation}
\author{H. \surname{Haider}}
\affiliation{Department of Physics, Aligarh Muslim University, Aligarh-202 002, India}
\author{I. \surname{Ruiz Simo}}
\affiliation{Departamento de F\'isica At\'omica Molecular y Nuclear, Universidad de Granada, E-18071 Granada, Spain}
\affiliation{Dipartimento di Fisica, Universit\`a degli studi di Trento, I-38123 Trento, Italy}
\author{M. Sajjad \surname{Athar}}
\email{sajathar@gmail.com}
\affiliation{Department of Physics, Aligarh Muslim University, Aligarh-202 002, India}
\begin{abstract}
We study nuclear medium effects and nonisoscalarity correction in the extraction 
of weak mixing angle sin$^2\theta_W$ using Paschos-Wolfenstein(PW) relation. The calculations are performed for the iron nucleus.
The nuclear medium effects like Fermi motion, 
binding, shadowing and antishadowing corrections and pion and rho meson cloud contributions have been taken into account. Calculations 
have been performed in the local density approximation using a relativistic nuclear spectral function which includes nucleon correlations. 
The results are discussed along with the experimental result inferred by the NuTeV Collaboration. These results may be useful for the ongoing MINER$\nu$A experiment as well as for the proposed NuSOnG experiment.
\end{abstract}
\maketitle
\section{Introduction}   
MINER$\nu$A~\cite{minerva1} is presently taking data using neutrinos from NuMI Lab., and their aim is to perform cross section 
measurements in the neutrino energy region of 1-20 GeV with different nuclear targets like helium, carbon, oxygen, iron and lead. 
 Among the various goals of the MINER$\nu$A experiment, one of them is to measure $\frac{d\sigma}{dxdQ^2}$ in the deep-inelastic scattering
 (DIS) region using these nuclear
targets, obtaining the ratio of the structure functions between the different target materials and also to study structure functions
in the DIS and transition region~\cite{minerva2,minerva3}. Neutrino Scattering On Glass (NuSOnG)~\cite{nusong} is another experiment 
proposed at Fermi lab to study the neutrino and antineutrino charged current deep-inelastic scattering events to 
precisely measure the structure functions $F_2^{\nu/{\bar\nu}}$, $xF_3^{\nu/{\bar\nu}}$, etc. 
 Furthermore, they plan to measure  $\sin^2 \theta_W$ from $\nu$-nucleon scattering using the Paschos-Wolfenstein(PW) relation~\cite{Paschos},
 a similar type of analysis was performed by
 the NuTeV group~\cite{Zeller:Thesis,Zeller:2001hh,McFarland}. Recently, we studied nuclear medium effects on 
 the electromagnetic structure function $F_{2}^{EM}$($x$,$Q^{2}$)~\cite{Sajjad1},  and the weak structure
 functions $F_{2}$($x$,$Q^{2}$) and $F_{3}$($x$,$Q^{2}$)~\cite{Sajjad2, Sajjad3}. For $F_{2}^{EM}$($x$,$Q^{2}$)~\cite{Sajjad1}, we found that our results are reasonably
   in good agreement with the recent results from JLab~\cite{Seely} as well as with some of the older experiments like SLAC~\cite{Gomez}. 
   In the case of $\nu(\bar\nu)$ deep inelastic scattering induced processes, the results were compared with the available data from 
   NuTeV, CDHSW and CHORUS experiments~\cite{Berge,Tzanov,Onengut} for the weak structure functions 
   $F_{2}$($x$,$Q^{2}$) and $xF_{3}$($x$,$Q^{2}$) in iron~\cite{Sajjad2} and lead~\cite{Sajjad3}, as well as with the results of the differential scattering cross section
   $\frac{1}{E}\frac{d\sigma}{dxdy}$. In this work, we study the effect of nuclear medium and nonisoscalarity correction in extracting $\sin^2 \theta_W$ using PW relation 
   taking iron as the nuclear target.
 
Paschos and Wolfenstein~\cite{Paschos} suggested that the ratio of neutral current to charged current neutrino interaction 
cross sections on nucleon targets may be used to measure $\sin^2 \theta_W$:
\begin{eqnarray} \label{ratio_cross1}
R_{PW}=\frac{\sigma(\nu_\mu~N \rightarrow \nu_\mu~X)~-~\sigma(\bar\nu_\mu~N \rightarrow \bar\nu_\mu~X)}{\sigma(\nu_\mu~N \rightarrow \mu^-~X)~-~\sigma(\bar\nu_\mu~N \rightarrow \mu^+~X)}=\frac{1}{2}~-~\sin^2 \theta_W
\end{eqnarray}
The above relation is valid when there is no contribution from heavy quarks, neglecting the charm quark mass, 
assuming the strange quark and anti-strange quark symmetry, considering no medium effect and no contribution from outside the standard model.
This equation is valid for both the total cross section $\sigma$ as well as differential cross sections 
$\frac{d^2\sigma}{dx\,dy}$, because the neutral current (NC) differential cross 
section can be expressed in terms of the charged current (CC) ones, and these cancel out in the quotient \cite{Bilenky}. 
For the total cross sections, the Paschos-Wolfenstein (PW) relationship is true under more general assumptions
(see for instance \cite{Paschos}, \cite{Bilenky} and \cite{McFarland:2003jw}).

The above relation is also valid for an isoscalar nuclear target(N=Z) for both the total cross sections and differential cross sections
and the above Eq.~(\ref{ratio_cross1}) may be written as 
\begin{eqnarray} \label{ratio_cross}
R_{PW}=\frac{\sigma(\nu_\mu~A \rightarrow \nu_\mu~X)~-~\sigma(\bar\nu_\mu~A \rightarrow \bar\nu_\mu~X)}{\sigma(\nu_\mu~A \rightarrow \mu^-~X)~-~\sigma(\bar\nu_\mu~A \rightarrow \mu^+~X)}
\end{eqnarray}
where $\sigma(\nu_\mu(\bar\nu_\mu)~A \rightarrow \nu_\mu(\bar\nu_\mu)~X)$ is the neutral current induced neutrino(antineutrino) 
cross section, $\sigma(\nu_\mu(\bar\nu_\mu)~A \rightarrow \mu^-(\mu^+)~X)$ is the charged current induced neutrino(antineutrino) cross section. The condition of pure isoscalarity includes the requirement of the cancellation
 of different strong interaction effects which also include the nuclear medium effects in the ratio of the neutral current to the charged current scattering cross sections. 

NuTeV Collaboration~\cite{Zeller:Thesis,Zeller:2001hh,McFarland} has measured
 the ratio {\it R} of neutral current to charged current total cross sections in iron, for which they took the ratio of charged current 
 antineutrino to neutrino cross sections i.e. 
$r=\frac{\sigma(\bar\nu_\mu~A \rightarrow \mu^+~X)}{\sigma(\nu_\mu~A \rightarrow \mu^-~X)}$ as $\frac{1}{2}$, 
and obtained the value for the weak mixing angle $\sin^2 \theta_W$ using Eq.~(\ref{ratio_cross}) as
\begin{eqnarray} \label{nutev}
R_{PW}&=&\frac{\frac{\sigma(\nu_\mu~A \rightarrow \nu_\mu~X)}{\sigma(\nu_\mu~A \rightarrow \mu^-~X)}-\frac{\sigma(\bar\nu_\mu~
A \rightarrow \bar\nu_\mu~X)}{\sigma(\bar\nu_\mu~A \rightarrow \mu^+~X)}
\frac{\sigma(\bar\nu_\mu~A \rightarrow \mu^+~X)}{\sigma(\nu_\mu~A \rightarrow \mu^-~X)}}{1-\frac{\sigma(\bar\nu_\mu~A \rightarrow \mu^+~X)}
{\sigma(\nu_\mu~A \rightarrow \mu^-~X)}}\nonumber \\
&=&\frac{R^\nu~-~r~R^{\bar\nu}}{1-r}
\end{eqnarray}
where $R^\nu_{exp}=0.3916\pm0.0007$ and $R^{\bar\nu}_{exp}=0.4050\pm0.0016$~\cite{Zeller:Thesis,Zeller:2001hh}. 
The reported value of $\sin^2 \theta_W$ is $0.2277\pm0.0004$~\cite{Zeller:Thesis,Zeller:2001hh,McFarland} which is
 3 standard deviations above the global fit of $sin^2 \theta_W=0.2227\pm0.0004$~\cite{Abbaneo:2001ix} and this is known as NuTeV anomaly.
 To resolve this anomaly, 
explanations within and outside the standard model of electroweak interactions have been looked for~\cite{Davidson:2001ji}-\cite{Thomas2}.

Paschos and Wolfenstein~\cite{Paschos} relation is valid for an isoscalar target while iron is a nonisoscalar target(N=30,Z=26), therefore,
nonisoscalar corrections are required. Furthermore, nuclear dynamics may also play 
an important role in the case of neutrino nucleus scattering. Various corrections made by the NuTeV Collaboration have been discussed in 
literature, but still the reported deviation could not be accounted for~\cite{McFarland}.
 Theoretically, Kulagin~\cite{Kulagin} has investigated the effect of nuclear medium on the PW ratio
 and pointed out that the shadowing effect being a low x and low $Q^2$ phenomenon is small in the $Q^2$ region of NuTeV experiment~\cite{Zeller2002}
 and observed the effects of Fermi motion and binding energy correction to be small but significant isoscalar correction.
Kumano\cite{Kumano:2002ra} in a phenomenological analysis pointed out the difference between the nuclear effects in the valence $u$ and $d$ 
quark distributions may be a reason for this anomaly. However, this effect is too small to explain the anomaly.
Recently Thomas~\cite{Thomas2} has discussed various possible corrections and concluded that charge symmetry violation and the isovector EMC
 effect together may explain this anomaly. 

In the present work, we used the results of our earlier study of nuclear medium and 
 nonisoscalarity correction on the weak structure functions and the differential scattering cross sections~\cite{Sajjad2, Sajjad3},
 on the extraction of $\sin^2 \theta_W$ using PW relation.  We have obtained the modified PW relation for a nonisoscalar nuclear target.  
  This study has been performed using a relativistic nucleon spectral function~\cite{FernandezdeCordoba:1991wf,FernandezdeCordoba:1995pt}, which
  is used to describe the momentum distribution of nucleons in the nucleus. We define everything
within a field-theoretical approach where nucleon propagators are written in
terms of this spectral function. The spectral function has been calculated
using the Lehmann's representation for the relativistic nucleon propagator and
nuclear many body theory is used for calculating it for an interacting Fermi
sea of nuclear matter. The local density approximation is then applied to
translate these results to finite nuclei~\cite{Sajjad1, Sajjad2, Sajjad3,Marco:1995vb,Sajjad}.
 The contributions
of the pion and rho meson clouds are taken into account in a many-body
field-theoretical approach which is based on
Refs.~\cite{Marco:1995vb,GarciaRecio:1994cn}. We have taken into account the target
mass correction (TMC) following Ref.~\cite{schienbein}, which has a
significant effect at low $Q^{2}$, and at moderate and high Bjorken $x$. To
take into account the shadowing effect, which is important at low $Q^{2}$ and
low $x$, and which modulates the contribution of pion and rho cloud
contributions, we have followed the works of Kulagin and Petti~\cite{Kulagin1,
Petti}. All the formalism is the same for neutral current scattering as done in the case of charged current neutrino/antineutrino induced reactions.
For the numerical calculations, parton distribution
 functions for the nucleons have been taken from the parametrization of CTEQ Collaboration (CTEQ6.6)~\cite{cteq}.

The paper is organised as follows. In Sec.~\ref{sec:formalism}, we present the formalism where we write the expression 
for the $\nu(\bar{\nu})$-nucleon differential scattering cross section in subsection-\ref{subsec:nun}, 
the expressions for $\nu(\bar{\nu})$-nucleus differential scattering cross section for the isoscalar
 as well as nonisoscalar nuclear targets are given in subsection-\ref{subsec:nua}. In subsection-\ref{subsec:hadronic_nuclear_tensor},
 we explicitly show the construction of the nuclear hadronic tensor for non-symmetric nuclear matter and in subsection-\ref{subsec:PW},
 the nuclear corrections to PW ratio are presented. In  
Sec.~\ref{sec:result}, we present and discuss the results of our calculations and finally our conclusions are summarised in Sec.~\ref{sec:concl}.
\section{Formalism}
\label{sec:formalism}
\subsection{Deep Inelastic Neutrino Nucleon Scattering}
\label{subsec:nun}
The cross section for the charged current(CC) neutrino interaction with a nucleon target i.e.
\begin{equation} 	\label{reaction}
\nu_l(k) + N(p) \rightarrow l^-(k^\prime) + X(p^\prime),~l=~e,~\mu,
\end{equation}
 is given by:
\begin{eqnarray} \label{csection}
\sigma&=&\frac{1}{v_{rel}}\frac{2m_\nu}{2E_{\nu}({\bf k})}\frac{2M}{2E({\bf p})}\int\frac{d^3 k^{\prime}}{(2\pi)^{3}}\frac{2m_l}{2E_{l}({\bf k}^{\prime})}
\prod_{i=1}^{N}\int\frac{d^3 p_{i}^{\prime}}{(2\pi)^{3}}\prod_{l\epsilon f}\left(\frac{2M^{\prime}_{l}}{2E^{\prime}_{l}}\right)\prod_{j\epsilon b}
\left(\frac{1}{2\omega^{\prime}_{j}}\right)\bar{\sum} \sum \left|T\right|^{2}  \nonumber \\
&&~~\times~~ (2\pi)^4 \delta^{4}\left(p+k-k^{\prime}-\sum_{i=1}^{N}p_{i}^{\prime}\right)
\end{eqnarray}
where $f$ stands for fermions and $b$ for bosons in the final state $X$. The index $i$ is split into $l$ and $j$ for fermions and bosons respectively, four momentum of the 
particles involved in the process are represented as $k$(incomig neutrino), $k^\prime$(outgoing lepton), $p$(target nucleon) and $p \prime$(jet of hadrons).

$T$ is the invariant matrix element for the above reaction and is,
written as
\begin{equation}	\label{Tinv}
- i T = \left( \frac{i G_F}{\sqrt{2}}\right) 
\bar{u}_l (k^\prime) 
\gamma^{\alpha} (1 -\gamma_5)u_{\nu} (k) \,
\left(\frac{m_W^2}{q^2-m_W^2}\right)\langle X | J_{\alpha} | N \rangle\,.
\end{equation}

After performing the phase space integration in Eq.~(\ref{csection}), the double differential scattering cross section evaluated for a nucleon target in its rest
frame is expressed as,
\begin{equation} 	\label{dif_cross}
\frac{d^2 \sigma_{\nu,\bar\nu}^N}{d \Omega' d E'} 
= \frac{{G_F}^2}{(2\pi)^2} \; \frac{|{\bf k}^\prime|}{|{\bf k}|} \;
\left(\frac{m_W^2}{q^2-m_W^2}\right)^2
L^{\alpha \beta}_{\nu, \bar\nu}
\; W_{\alpha \beta}^{N}\,,
\end{equation}
where $G_F$ is the Fermi coupling constant, $m_W$ is the mass of the W boson, $l(=e,\,\mu)$ is a lepton, $q=k-k^\prime$ is the four momentum transfer and $\Omega', E'$ refer to the outgoing lepton.
$N$ is a nucleon, $X$ is a jet of n hadrons consisting of fermions(f) and bosons(b) in the final state.

The leptonic tensor for antineutrino(neutrino) scattering $L^{\alpha \beta}$ is given by
\begin{equation} 	\label{dif_cross2}
L^{\alpha \beta}=k^{\alpha}k'^{\beta}+k^{\beta}k'^{\alpha}
-k.k^\prime g^{\alpha \beta} \pm i \epsilon^{\alpha \beta \rho \sigma} k_{\rho} 
k'_{\sigma}\,,
\end{equation}
and the hadronic tensor $W^{N}_{\alpha \beta}$ is defined as
\begin{eqnarray}\label{hadten}
W^{N}_{\alpha \beta}&=&\frac{1}{2\pi} \bar{\sum_{s_N}} \;
\sum_X \; \sum_{s_i} \prod^n_{i = 1}
\; \int \frac{d^3 p'_i}{(2 \pi)^3} 
\prod_{l \epsilon f} \;
\left(
\frac{2 M'_l}{2 E'_l}
\right) \;
\prod_{j \epsilon b} \;
\left(
\frac{1}{2 \omega'_j}
\right)
\langle X | J_{\alpha} | N \rangle 
\langle X | J_{\beta} | N \rangle^*  \nonumber \\
&&~~\times~~(2 \pi)^4 \delta^4 (p + q - \sum^n_{i = 1} p'_i)\,,
\end{eqnarray}
where $s_N$ the spin of the
nucleon and $s_i$ the spin of the fermions in $X$. In the case of antineutrino $\langle X | J_{\alpha} | N \rangle$
is replaced by $\langle X | J^{\dagger}_{\alpha} | N \rangle$.

The most general form of the hadronic tensor $W^{N}_{\alpha \beta}$ is expressed as:
\begin{eqnarray} \label{had_ten}
W^{N}_{\alpha \beta} =&& 
\left( \frac{q_{\alpha} q_{\beta}}{q^2} - g_{\alpha \beta} \right) \;
W_1^{\nu (\bar{\nu})}
+ \frac{1}{M^2}\left( p_{\alpha} - \frac{p . q}{q^2} \; q_{\alpha} \right)
\left( p_{\beta} - \frac{p . q}{q^2} \; q_{\beta} \right)
W_2^{\nu (\bar{\nu})}-\frac{i}{2M^2} \epsilon_{\alpha \beta \rho \sigma} p^{\rho} q^{\sigma}
W_3^{\nu (\bar{\nu})} + \nonumber\\
&&
\frac{1}{M^2} q_{\alpha} q_{\beta}
W_4^{\nu (\bar{\nu})}
+\frac{1}{M^2} (p_{\alpha} q_{\beta} + q_{\alpha} p_{\beta})
W_5^{\nu (\bar{\nu})}
+ \frac{i}{M^2} (p_{\alpha} q_{\beta} - q_{\alpha} p_{\beta})
W_6^{\nu (\bar{\nu})}\,,
\end{eqnarray}
where $M$ is the nucleon mass and $W_i^{N}$ are the structure functions, which depend
on the scalars $q^2$ and $p.q$. The terms depending on $W_4$, $W_5$
and $W_6$ in Eq.~(\ref{had_ten}) do not contribute to the cross
section in Eq.~(\ref{dif_cross}) in the limit of lepton mass $m_l \rightarrow 0$.

In terms of the Bjorken variables $x$ and y defined as
\begin{eqnarray}	\label{Bj_var}
x=\frac{Q^2}{2M\nu}, \quad y=\frac{\nu}{E_{\nu}}, \quad Q^2=-q^2, \quad \nu=\frac{p.q}{M} ,
\end{eqnarray}
$W_i^{N}$ are expressed in terms of dimensionless
structure functions $F_i^{\nu, \bar\nu} (x,Q^2)$   
\begin{eqnarray}\label{relation}
F_1^{\nu(\bar{\nu})}(x, Q^2)&=&M W_1^{\nu(\bar{\nu})}(\nu, Q^2)  \nonumber\\
F_2^{\nu(\bar{\nu})}(x, Q^2)&=&\nu W_2^{\nu(\bar{\nu})}(\nu, Q^2) \nonumber\\ 
F_3^{\nu(\bar{\nu})}(x, Q^2)&=&\nu W_3^{\nu(\bar{\nu})}(\nu, Q^2).
\end{eqnarray}
The expression of the differential cross section, for deep inelastic scattering (DIS) of neutrino with a nucleon target induced by charged current reaction 
is now given by:
\begin{footnotesize}
\begin{eqnarray} \label{cross_section}
\frac{d^2\sigma^{\nu(\bar{\nu})}}{dx\ dy} &=& \frac{G_F^2 M
E_{\nu}}{\pi(1+Q^2/M_W^2)^2}\Biggl((y^2 x + \frac{m_l^2 y}{2 E_{\nu} M})
F_1(x,Q^2) +\left[ (1-\frac{m_l^2}{4 E_{\nu}^2})
-(1+\frac{M x}{2 E_{\nu}}) y\right]F_2(x,Q^2) \\ \nonumber
&
\pm&
\left[x y (1-\frac{y}{2})\right.
-\left.\frac{m_l^2 y}{4 E_{\nu} M}\right]
F_3(x,Q^2) \Biggr)\
\end{eqnarray}
\end{footnotesize}
In the above equation +sign(-sign) in the coefficient with $F_3$ is for neutrino(antineutrino). 
$F_1$ and $F_2$ are related by the Callan-Gross relation, leading to only
two independent structure functions $F_2$ and $F_3$. For l=e,$\mu$ we take $m_l=0$ and assume $Q^2 << M_W^2$. 

The nucleon structure functions are determined in terms of parton distribution functions for quarks and anti-quarks given by
\begin{eqnarray}\label{f2f3}
F^{\nu p}_{2}& = & 2 x [d(x) + s(x) + \bar{u}(x) +\bar{c}(x)]~~~~F^{\bar{\nu} p}_{2} =  2 x [u(x) + c(x) + \bar{d}(x) +\bar{s}(x)]\nonumber\\ 
F^{\nu n}_{2} & = & 2 x [u(x) + s(x) + \bar{d}(x) +\bar{c}(x)]~~~~F^{\bar{\nu} n}_{2}  =  2 x [d(x) + c(x) + \bar{u}(x) +\bar{s}(x)]\nonumber\\
xF^{\nu p}_{3}&=& 2x[d(x) + s(x) -{\bar u}(x) - {\bar c}(x)]~~~~xF^{\nu n}_{3} = 2x[u(x) + s(x) -{\bar d}(x) - {\bar c}(x)]\nonumber\\
xF^{{\bar \nu} p}_{3}&= &2x[u(x) + c(x) -{\bar d}(x) - {\bar s}(x)]~~~~xF^{{\bar \nu} n}_{3} = 2x[d(x) + c(x) -{\bar u}(x) - {\bar s}(x)]
\end{eqnarray}

For the neutral current(NC) induced reaction
\begin{eqnarray}
\nu_l(\bar{\nu}_l)(k) + N(p) \rightarrow \nu_l(k^\prime) + X(p^\prime),~l=~e,~\mu,~\tau
\end{eqnarray}
the expression of the cross section(\ref{cross_section}) modifies by changing $M_W \rightarrow M_Z$, the mass of $Z^0$ boson and the corresponding NC structure functions are given by
\begin{eqnarray}
F_2^{NC}(\nu p,\bar{\nu} p)& = & 2 x ((u_L^2+u_R^2) [u + c + \bar{u} +\bar{c}]+(d_L^2+d_R^2) [d + s + \bar{d} +\bar{s}])\\\nonumber
x F_3^{NC}(\nu p,\bar{\nu} p) & = &  2 x ((u_L^2-u_R^2) [u + c -\bar{u} -\bar{c}]+(d_L^2-d_R^2) [d + s -\bar{d} -\bar{s}])\nonumber
\end{eqnarray}
for the proton target and  
\begin{eqnarray}
F_2^{NC}(\nu n, \bar{\nu} n) & = & 2 x ((u_L^2+u_R^2) [d + c + \bar{d} +\bar{c}]+(d_L^2+d_R^2) [u + s + \bar{u} +\bar{s}])\\\nonumber
x F_3^{NC}(\nu n, \bar{\nu} n) &  = & 2 x ((u_L^2-u_R^2) [d + c -\bar{d} -\bar{c}]+(d_L^2-d_R^2) [u + s -\bar{u} - \bar{s}]),\nonumber
\end{eqnarray}  
for the neutron target. Here $u_{L} = \frac{1}{2} -\frac{2}{3} \sin^2 \theta_W, \,u_R = -\frac{2}{3} \sin^2 \theta_W$ and 
$d_L = - \frac{1}{2} + \frac{1}{3} \sin^2 \theta_W, \,d_R = \frac{1}{3} \sin^2 \theta_W$.
\subsection{Deep Inelastic Neutrino Nucleus Scattering}
\label{subsec:nua}
When the reaction given by Eq.~(\ref{reaction}) takes place in a nucleus, several nuclear effects have to be considered. 
One may categorize these medium effects into two parts, a kinematic effect which arises because the struck nucleon is not at rest 
but is moving with a Fermi momentum in the rest frame of the nucleus and the other one is a dynamic effect
which arises due to the strong interaction of the initial nucleon in the nuclear medium. 
For details please see the discussion given in Refs.~\cite{Marco:1995vb,Sajjad}.

The expression for the differential scattering cross section for a nuclear target A is similar to Eq.~(\ref{dif_cross}) and is given by:
\begin{equation}
\frac{d\,\sigma^{\nu(\bar{\nu})A}_{CC}}{dE'\;d\Omega'}=\frac{G^2_F}{(2\pi)^2}\cdot\frac{\left|\vec{k}'\right|}{\left|\vec{k}\right|}\cdot\left(\frac{m^2_W}{q^2-m^2_W}\right)^2\cdot L^{\alpha\beta}_{\nu,\bar{\nu}}W^{\nu(\bar{\nu})A}_{\alpha\beta}
\label{eq:diff_cross_section_nu_antinu_nucleus}
\end{equation}
where $L^{\alpha\beta}_{\nu,\bar{\nu}}$ is given by Eq.~(\ref{dif_cross2}) and $W^{\nu(\bar{\nu})A}_{\alpha\beta}$, the nuclear 
hadronic tensor, is given by:
\begin{eqnarray}
W^{\nu(\bar{\nu})A}_{\alpha\beta}&=&\left(\frac{q_\alpha q_\beta}{q^2}-g_{\alpha\beta}\right)W^{\nu(\bar{\nu})A}_1(P_A,q)\nonumber\\
&+&\frac{1}{M^2_A}\left(P_{A\alpha}-\frac{P_A\cdot q}{q^2}q_\alpha\right)\left(P_{A\beta}-\frac{P_A\cdot q}{q^2}q_\beta\right)W^{\nu(\bar{\nu})A}_2(P_A,q)\nonumber\\
&-&\frac{\imath}{2M^2_A}\;\epsilon_{\alpha\beta\rho\sigma}\; P^\rho_Aq^\sigma W^{\nu(\bar{\nu})A}_3(P_A,q)\label{eq:nuclear_hadronic_tensor}
\end{eqnarray}
where $P_A$ is the momentum of the nucleus A. 

In the local density approximation, the nuclear hadronic tensor $W^{\nu(\bar{\nu})A}_{\alpha\beta}$ can be written as a convolution of the nucleonic hadronic tensor
 with the hole spectral function. For symmetric nuclear matter, this would be~\cite{Sajjad2}:
\begin{equation}\label{iso}
W^{\nu(\bar{\nu})A}_{\alpha\beta}=4\int d^3r\int\frac{d^3p}{(2\pi)^3}\frac{M}{E(\mathbf{p})}\int^{\mu}_{-\infty}dp^0\; S_h(p^0,\mathbf{p},k_F(\vec{r}))~ W^{\nu(\bar{\nu})N}_{\alpha\beta},
\label{eq:convolution_hadronic_tensor_symm_nuclear_matter}
\end{equation}
where $k_F(\vec{r})$ is the Fermi momentum for symmetric nuclear matter which depends on the density of nucleons in the nucleus, i.e. $k_F(\vec{r})=\left(\frac{3\pi^2}{2}\rho(\vec{r})\right)^{1/3}$. 
$S_h(p^0,\mathbf{p},k_F(\vec{r}))$ is the hole spectral function, $\mu$ is the chemical 
potential and have been taken from Ref.~\cite{FernandezdeCordoba:1991wf}. 
$W^{\nu(\bar{\nu})N}$ is the hadronic tensor for the free nucleon target given by Eq.~(\ref{had_ten}). 
M and $E(\mathbf{p})$ are respectively the mass and energy of the nucleon.

The natural extension of the above expression for taking into account the non-symmetric nature of the target nucleus would be 
to consider separate distributions of Fermi seas for protons and neutrons. The expression for which is given by~\cite{Sajjad3}:
\begin{eqnarray}\label{eq:convolution_hadronic_tensor_nonsymm_nuclear_matter}
W^{\nu(\bar{\nu})A}_{\alpha\beta}&=&2\int d^3r\int\frac{d^3p}{(2\pi)^3}\frac{M}{E(\mathbf{p})}\int^{\mu_p}_{-\infty}dp^0\; S^{proton}_{h}
(p^0,\mathbf{p},k_{F,p})\cdot W^{\nu(\bar{\nu})p}_{\alpha\beta} \nonumber\\
&+&~2\int d^3r\int\frac{d^3p}{(2\pi)^3}\frac{M}{E(\mathbf{p})}\int^{\mu_n}_{-\infty}dp^0\; S^{neutron}_{h}(p^0,\mathbf{p},k_{F,n})\cdot W^{\nu(\bar{\nu})n}_{\alpha\beta}
\end{eqnarray}
where the factor $2$ in front of the integral accounts for the two degrees of freedom of the spin of the nucleons. 
In the above equation, $S^{proton}_{h}$ and $S^{neutron}_{h}$ are the two different spectral functions and normalized respectively to the number of protons or neutrons in the nuclear target. 
$k_{F,p}=(3\pi^2\rho_p)^{1/3}$ ($k_{F,n}=(3\pi^2\rho_n)^{1/3}$) is the Fermi momentum of proton (neutron). 
For the proton and neutron densities in iron, we have used two-parameter Fermi density distribution 
and the density parameters are taken from Ref.~\cite{Vries}.
\subsection{Construction of the nuclear hadronic tensor for non-symmetric nuclear matter}
\label{subsec:hadronic_nuclear_tensor}
The natural invariant quantities for deep inelastic scattering (DIS) of neutrinos with nuclei are:
\begin{eqnarray}
 x_A=\frac{Q^2}{2P\cdot q};\ \ \nu_A=\frac{P\cdot q}{M_A};\ \ y_A=\frac{P\cdot q}{P\cdot k}
\label{eq:invariant_quantities}
\end{eqnarray}
where $x_A$ is the Bjorken variable in the nucleus and $x_A\in\left[0,1\right]$; $y_A$ is the inelasticity. These two variables are related to the nucleonic ones via:
\begin{eqnarray}
x_A&=&\frac{x}{A};~~y_A=\frac{q^0}{E_\nu}=y\label{eq:yA_y}
\end{eqnarray}
where $x$ and $y$ are defined in Eq.~(\ref{Bj_var}). We can see that $x\in\left[0,A\right]$, although for $x>1$ the nuclear structure functions are negligible. The variable $y_A$ varies between the following limits:
\begin{equation}
0\leq y_A\leq\frac{1}{1+\frac{M_Ax_A}{2E_\nu}}\approx\frac{1}{1+\frac{Mx}{2E_\nu}},
\label{eq:limits_of_yA}
\end{equation}
so, for sufficient high neutrino energy we have $0\leq y_A\leq1$.\\
If we express the differential cross section with respect to these variables ($x_A,y_A$), we obtain the following expression in terms of the nuclear structure functions:
\begin{eqnarray}\label{cc_xsection}
\frac{d^2\sigma^{\nu(\bar{\nu})A}_{CC}}{dx_A\;dy_A}&=&\frac{G^2_FM_AE_\nu}{\pi}\left(\frac{m^2_W}{Q^2+m^2_W}\right)^2\Bigg(y^2_Ax_AF^{\nu(\bar{\nu})A}_{1}\nonumber\\
&+&\left\{1-y_A-\frac{M_Ax_Ay_A}{2E_\nu}\right\}F^{\nu(\bar{\nu})A}_{2}\pm x_Ay_A\left(1-\frac{y_A}{2}\right)F^{\nu(\bar{\nu})A}_{3}\Bigg)\label{eq:CC_cross_section_xA_yA}
\end{eqnarray}
For the neutral current induced neutrino interaction, the form of the differential cross section is the same as for the charged 
current induced process but with the following changes:
\begin{eqnarray}\label{nc_changes}
\frac{d^2\sigma^{\nu(\bar{\nu})A}_{CC}}{dx_A\;dy_A}&\longrightarrow&\frac{d^2\sigma^{\nu(\bar{\nu})A}_{NC}}{dx_A\;dy_A}\nonumber\\
m_W&\longrightarrow& m_Z \nonumber\\
F^{CC}_{i}&\longrightarrow& F^{NC}_{i}
\end{eqnarray}
First we look at the denominator of the Paschos-Wolfenstein relationship, an expression similar to Eq.~(\ref{ratio_cross}), but in 
terms of the differential scattering cross section, for which we subtract the charged current 
antineutrino-nucleus cross section from the charged current neutrino-nucleus cross section and obtain the expression as:
\begin{footnotesize}
\begin{eqnarray}\label{difference}
\frac{d^2\sigma^{\nu A}_{CC}}{dx_A\;dy_A}-\frac{d^2\sigma^{\bar{\nu}A}_{CC}}{dx_A\;dy_A}&=&\frac{G^2_FM_AE_\nu}{\pi} 
\Bigg[y^2_Ax_A\left(F^{\nu A}_1-F^{\bar{\nu}A}_1\right)  \nonumber\\
&&+\left\{1-y_A-\frac{M_Ax_Ay_A}{2E_\nu}\right\}\left(F^{\nu A}_2-F^{\bar{\nu}A}_2\right)
+x_Ay_A\left(1-\frac{y_A}{2}\right)\left(F^{\nu A}_3+F^{\bar{\nu}A}_3\right)\Bigg]~~\label{eq:denominator_PW}
\end{eqnarray}
\end{footnotesize}
Since $Q^2 < < M_W^2$ in the present study, we have neglected the W-boson propagator term. 

We need to relate the nuclear structure functions $F_i^A$ to the nucleon ones via an integral with the spectral function. Therefore, we introduce the following notation to avoid writing every time the integration symbols.
For example, we may rewrite Eq.~(\ref{eq:convolution_hadronic_tensor_nonsymm_nuclear_matter}) with the following notation:
\begin{equation}
W^{\nu(\bar{\nu})A}_{\alpha\beta}=\left\langle W^{\nu(\bar{\nu})p}_{\alpha\beta}\right\rangle_{S^{proton}_h}+\left\langle W^{\nu(\bar{\nu})n}_{\alpha\beta}\right\rangle_{S^{neutron}_h}
\label{eq:notation_hadronic_tensor}
\end{equation}
where $\left\langle W^{\nu(\bar{\nu})p}_{\alpha\beta}\right\rangle_{S^{proton}_h}$ stands for proton:

\begin{equation}\label{eq29}
\left\langle W^{\nu(\bar{\nu})p}_{\alpha\beta}\right\rangle_{S^{proton}_h}=2\int d^3r\int\frac{d^3p}{(2\pi)^3}\frac{M}{E(\mathbf{p})}\int^{\mu_p}_{-\infty}dp^0\;S^{proton}_h(p^0,\mathbf{p},k_{F,p})W^{\nu(\bar{\nu})p}_{\alpha\beta}
\label{eq:explanation_notation}
\end{equation}
and for the neutron the expression is the same when the indices for the proton are replaced by the neutron indices. 

Taking the three momentum transfer $\vec{q}$ along the Z-axis, i.e, $q^\mu=(q^0,0,0,q_z)$, and writing the 
xx-component of the nuclear hadronic tensor (Eq.~\ref{eq:nuclear_hadronic_tensor}), we get it in terms of the nuclear structure function $F_1$ i.e.
\begin{equation}
W^{\nu(\bar{\nu})A}_{xx}=W^{\nu(\bar{\nu})A}_{1}=\frac{F^{\nu(\bar{\nu})A}_1(x_A)}{M_A}=W^{\nu(\bar{\nu})A}_{yy}
\label{eq:xx_component}
\end{equation}
Similarly, if we take the xx-components of the nucleonic hadronic tensor given by Eq.~(\ref{had_ten}) and remembering 
the fact that nucleons in the nucleus are not at rest, the xx-component of the nucleonic hadronic tensor is not related only to the nucleon structure function 
$F_1$, but it is a mixture of $F_1$ and $F_2$ components like the expression written below:
\begin{equation}
W^{\nu(\bar{\nu})N}_{xx}=W^{\nu(\bar{\nu})N}_{1}+\frac{p^2_x}{M^2}W^{\nu(\bar{\nu})N}_{2}=\frac{F^{\nu(\bar{\nu})N}_1(x_N)}{M}+\frac{p^2_x}{M^2}\frac{F^{\nu(\bar{\nu})N}_2(x_N)}{\nu_N}
\label{eq:xx_component_nucleonic_tensor}
\end{equation}
where $N\equiv p,n$; $x_N\equiv\frac{Q^2}{2p\cdot q}=\frac{Q^2}{2(p^0q^0-p_zq_z)}$ and $\nu_N=\frac{p\cdot q}{M}=\frac{p^0q^0-p_zq_z}{M}$.

Using Eqs. (\ref{eq:xx_component}) and (\ref{eq:xx_component_nucleonic_tensor}) we may write:
\begin{equation}\label{eq32}
\frac{F^{\nu(\bar{\nu})A}_1(x_A)}{M_A}=\left\langle \frac{F^{\nu(\bar{\nu})p}_1(x_N)}{M}+\frac{p^2_x}{M}\frac{F^{\nu(\bar{\nu})p}_2(x_N)}{p\cdot q}\right\rangle_{S^p_h}+\left\langle \frac{F^{\nu(\bar{\nu})n}_1(x_N)}{M}+\frac{p^2_x}{M}\frac{F^{\nu(\bar{\nu})n}_2(x_N)}{p\cdot q}\right\rangle_{S^n_h}
\label{eq:nuclear_structure_function_F1}
\end{equation}
The difference $F^{\nu A}_1-F^{\bar{\nu}A}_1$ that appears in Eq.~(\ref{eq:denominator_PW}) may then be written as:
\begin{eqnarray}\label{eq:difference_F1}
\frac{F^{\nu A}_1(x_A)}{M_A}-\frac{F^{\bar{\nu}A}_1(x_A)}{M_A}&=&\left\langle \frac{1}{M}\left(F^{\nu p}_1(x_N)-F^{\bar{\nu} p}_1(x_N)\right)+
\frac{p^2_x}{M(p\cdot q)}\left(F^{\nu p}_2(x_N)-F^{\bar{\nu} p}_2(x_N)\right)\right\rangle_{S^p_h} \nonumber\\
&+&\left\langle \frac{1}{M}\left(F^{\nu n}_1(x_N)-F^{\bar{\nu} n}_1(x_N)\right)+\frac{p^2_x}{M(p\cdot q)}\left(F^{\nu n}_2(x_N)-
F^{\bar{\nu} n}_2(x_N)\right)\right\rangle_{S^n_h}~~~~\;\;
\end{eqnarray}
where
\begin{eqnarray*}
F^{\nu p}_1-F^{\bar{\nu}p}_1&=&d_v-u_v\\ \nonumber
F^{\nu p}_2-F^{\bar{\nu}p}_2&=&2x_N\left(d_v-u_v\right)\\  \nonumber
F^{\nu n}_1-F^{\bar{\nu}n}_1&=&-d_v+u_v=-\left(F^{\nu p}_1-F^{\bar{\nu}p}_1\right)\label{eq:antisymmetry_F1}\\ \nonumber
F^{\nu n}_2-F^{\bar{\nu}n}_2&=&2x_N\left(-d_v+u_v\right)=-\left(F^{\nu p}_2-F^{\bar{\nu}p}_2\right)\label{eq:antisymmetry_F2}
\end{eqnarray*}
Here $u_v$ and $d_v$ are the valence PDFs and we are working in the so-called \textit{up and down quarks approximation}, where we neglect strange and charm quarks contributions.

In the case of symmetric nuclear matter, we may relate the Fermi momentum with the baryon density 
via $k_F^3=\frac{3\pi^2}{2}\rho$, where $\rho$ is the baryon density. 
For a non-symmetric nuclear matter, we have different densities for protons and neutrons and 
corresponding to those, we also have different Fermi momenta for protons and neutrons. These are related by
\begin{equation}
k^3_{F,p}=3\pi^2\rho_p(r){\bf ;}\;\;\;k^3_{F,n}=3\pi^2\rho_n(r)
\end{equation}
Instead of discussing in terms of neutron number(N) and proton number(Z) as independent variables, we define two independent 
variables A=N+Z and their difference $\delta$=N-Z such that:
\begin{eqnarray}
N&=&\frac{A}{2}+\frac{\delta}{2}{\bf ;}~~Z=\frac{A}{2}-\frac{\delta}{2}
\end{eqnarray}
Dividing the above equations by the nuclear volume $V$, we obtain the densities of neutrons and protons:
\begin{eqnarray}
\rho_n&=&\frac{\rho}{2}+\frac{\delta}{2V}{\bf ;}\;\;\;\rho_p=\frac{\rho}{2}-\frac{\delta}{2V}
\end{eqnarray}
where $\rho=\rho_p+\rho_n$ and $\frac{\delta}{V}=\frac{N-Z}{V}=\rho_n-\rho_p$. Replacing the densities for neutrons and protons
by their corresponding Fermi momenta one has $k_{F,p/n}$ in terms of $k_F$, $\delta$ and V i.e.
\begin{eqnarray}
k^3_{F,p}&=&k^3_F-\frac{3\pi^2}{2}\frac{\delta}{V}{\bf ;}\;\;\;k^3_{F,n}=k^3_F+\frac{3\pi^2}{2}\frac{\delta}{V}.
\end{eqnarray}
A non-zero value of $\delta$ would imply that we are looking for the deviations from the isoscalarity. 
For $k_{F,p}$ and $k_{F,n}$, we are going to perform an expansion in powers of the parameter $\frac{\delta}{V}$, and retaining the first order term only with the assumption
that the higher orders would be negligible. For instance, the expansion for the Fermi momentum of the proton would be:
\begin{eqnarray}
k_{F,p}&=&\left(k^3_F-\frac{3\pi^2}{2}\frac{\delta}{V}\right)^{1/3}= k_F-\frac{\pi^2}{2k^2_F}\frac{\delta}{V}+\mathcal{O}\left(\frac{\delta}{V}\right)^2
\end{eqnarray}
and the proton's spectral function may be written as:
\begin{equation}
S_h^{proton}(p^0,\mathbf{p},k_{F,p})\simeq S_h^p(p^0,\mathbf{p},k_F-\frac{\pi^2}{2k^2_F}\frac{\delta}{V})\simeq S_h^p(p^0,\mathbf{p},k_F)+
\frac{\partial S_h^p(p^0,\mathbf{p},k)}{\partial k}\Big|_{k=k_F}\left(-\frac{\pi^2}{2k^2_F}\frac{\delta}{V}\right).
\label{eq:expansion_hole_spectral_function}
\end{equation}
With a  change of $\delta\rightarrow-\delta$ one gets the neutron spectral function.

Using Eq.(\ref{eq32}), one may write $F_1^{\nu A}-F_1^{\bar{\nu}A}$ as 
\begin{eqnarray}
F_1^{\nu A}-F_1^{\bar{\nu}A}&=&M_A\left\langle\frac{F_1^{\nu p}-F_1^{\bar{\nu}p}}{M}+\frac{p^2_x}{M(p\cdot q)}\left(F_2^{\nu p}-
F_2^{\bar{\nu}p}\right)\right\rangle_{S_h^p} \nonumber\\
&+&M_A\left\langle\frac{F_1^{\nu n}-F_1^{\bar{\nu}n}}{M}+\frac{p^2_x}{M(p\cdot q)}\left(F_2^{\nu n}-F_2^{\bar{\nu}n}\right)\right\rangle_{S_h^n}.
\end{eqnarray}

If we look at the convolution with the proton spectral function given by Eq.(\ref{eq29}), we may write the first term of the right hand side as:
\begin{eqnarray}
2M_A\int d^3r\int\frac{d^3p}{(2\pi)^3}\frac{M}{E(\mathbf{p})}\int_{-\infty}^{\mu_p}dp^0\, S_h^p(p^0,\mathbf{p},k_{F,p})
\left(\frac{F_1^{\nu p}-F_1^{\bar{\nu}p}}{M}+\frac{p^2_x}{M(p\cdot q)}\left(F_2^{\nu p}-F_2^{\bar{\nu}p}\right)\right)~\label{eq:expansion_convolution_F1}~~
\end{eqnarray}
where $\mu_p=M+\frac{k^2_{F,p}}{2M}$ is the chemical potential, which must be expanded around the isoscalarity condition. $\mu_p$ may be written as:
\begin{equation}
\mu_p=M+\frac{k^2_{F,p}}{2M}\simeq M+\frac{1}{2M}\left(k^2_F-\frac{\pi^2}{k_F}\frac{\delta}{V}\right)
\end{equation}
Inserting Eq.~(\ref{eq:expansion_hole_spectral_function}) in Eq.~(\ref{eq:expansion_convolution_F1}), we obtain the following
 expression with the  spectral function:
\begin{eqnarray}
&&2M_A\int d^3r\int\frac{d^3p}{(2\pi)^3}\frac{M}{E(\mathbf{p})}\int_{-\infty}^{M+\frac{1}{2M}
\left(k^2_F-\frac{\pi^2}{k_F}\frac{\delta}{V}\right)}dp^0\,
\left\{S_h^p(p^0,\mathbf{p},k_F)\right.~\nonumber\\
&&~~~~\left.+\frac{\partial S_h^p(p^0,\mathbf{p},k)}{\partial k}\Big|_{k=k_F}\left(-\frac{\pi^2}{2k^2_F}
\frac{\delta}{V}\right)\right\}G^{\nu p-\bar{\nu}p}(\mathbf{p},x_N)\label{eq:expansion_convolution_F1_bis}
\end{eqnarray}
where
\begin{equation}
G^{\nu p-\bar{\nu}p}(\mathbf{p},x_N)=\left(\frac{F_1^{\nu p}-F_1^{\bar{\nu}p}}{M}+\frac{p^2_x}{M(p\cdot q)}\left(F_2^{\nu p}-F_2^{\bar{\nu}p}\right)\right)\label{eq:definition_Gnuproton}
\end{equation}
Integrating the first term of Eq.~(\ref{eq:expansion_convolution_F1_bis}), over the variable $p^0$:
\begin{eqnarray}
\int_{-\infty}^{M+\frac{k^2_F}{2M}-\frac{\pi^2}{2Mk_F}\frac{\delta}{V}}&&dp^0\,S_h^p(p^0,\mathbf{p},k_F)G^{\nu p-\bar{\nu}p}(p^0,\mathbf{p},x_N)\nonumber\\
&=&\int_{-\infty}^\mu dp^0S_h^p(p^0,\mathbf{p},k_F)G^{\nu p-\bar{\nu}p}(p^0,\mathbf{p},x_N)\nonumber\\
&+&\int_\mu^{\mu-\frac{\pi^2}{2Mk_F}\frac{\delta}{V}}dp^0\,S_h^p(p^0,\mathbf{p},k_F)G^{\nu p-\bar{\nu}p}(p^0,\mathbf{p},x_N)\nonumber\\
&\simeq&\int_{-\infty}^\mu dp^0S_h^p(p^0,\mathbf{p},k_F)G^{\nu p-\bar{\nu}p}(p^0,\mathbf{p},x_N)~~ \nonumber\\
&&~~+~~S_h^p(\mu,\mathbf{p},k_F)~G^{\nu p-\bar{\nu}p}(p^0=\mu,\mathbf{p},x_N)~\left(-\frac{\pi^2}{2Mk_F}\frac{\delta}{V}\right)~\label{eq:first_term_integral}
\end{eqnarray}
where $\mu=M+\frac{k^2_F}{2M}$ and we have used the following property:
\begin{equation}
\int_\mu^{\mu+\delta x}dy\,f(y)\simeq f(\mu)\times\delta x
\end{equation}
for $\delta x\to0$.

Let us analyze what we have obtained from the first term of the integral in Eq.(\ref{eq:expansion_convolution_F1_bis}).
We have obtained an isoscalar term (the first term in Eq.~\ref{eq:first_term_integral}), which still has to be integrated over
 $p^0$, plus a correction proportional to $\frac{\delta}{V}$. 
This correction would be zero for an isoscalar target.

In Eq.(\ref{eq:expansion_convolution_F1_bis}), we still have another integral 
(the one that goes with the partial derivative) which is already of order $\mathcal{O}\left(\frac{\delta}{V}\right)$. 
Therefore, we will only have to calculate the contribution coming from the limits of integration in $p^0$ at order $\mathcal{O}(1)$. 
Indeed, we have:
\begin{eqnarray}
\int_{-\infty}^{\mu-\frac{\pi^2}{2Mk_F}\frac{\delta}{V}}&&dp^0\frac{\partial S_h^p(p^0,\mathbf{p},k)}{\partial k}\Big|_{k=k_F}
\left(-\frac{\pi^2}{2k^2_F}\frac{\delta}{V}\right)G^{\nu p-\bar{\nu}p}(p^0,\mathbf{p},x_N)\nonumber\\
&=&\left(-\frac{\pi^2}{2k^2_F}\frac{\delta}{V}\right)\int_{-\infty}^\mu dp^0\frac{\partial S_h^p(p^0,\mathbf{p},k)}{\partial k}\Big|_{k=k_F}G^{\nu p-\bar{\nu}p}(p^0,\mathbf{p},x_N)+\mathcal{O}\left(\frac{\delta}{V}\right)^2
\end{eqnarray}
Therefore, Eq.(\ref{eq:expansion_convolution_F1}) may be written as
\begin{eqnarray}
&2&M_A\int d^3r\int\frac{d^3p}{(2\pi)^3}\frac{M}{E(\mathbf{p})}\Bigg\lbrace\int_{-\infty}^{\mu}dp^0\,S_h^p(p^0,
\mathbf{p},k_F)G^{\nu p-\bar{\nu}p}(p^0,\mathbf{p},x_N)\nonumber\\
&-&\frac{\pi^2}{2Mk_F(r)}\frac{\delta}{V}S_h^p(\mu,
\mathbf{p},k_F)G^{\nu p-\bar{\nu}p}(\mu,\mathbf{p},x_N)-\frac{\pi^2}{2k^2_F}\frac{\delta}{V}\int_{-\infty}^\mu dp^0\frac{\partial S_h^p(p^0,\mathbf{p},k)}{\partial k}\Big|_{k=k_F}
G^{\nu p-\bar{\nu}p}(p^0,\mathbf{p},x_N)\Bigg\rbrace~~\label{eq:expansion_convolution_F1_proton}~~~
\end{eqnarray}
This result is only for the convolution with the proton-hole spectral function. Similarly there will be a corresponding term for the
neutron-hole spectral function. But the changes are minimal. They reduce to change $p$ (protons) by $n$ (neutrons) in the structure functions 
and $\delta\to-\delta$. With these changes, we may write the equivalent expression to the above one for the convolution with the neutron 
spectral function as:
\begin{eqnarray}
&2&M_A\int d^3r\int\frac{d^3p}{(2\pi)^3}\frac{M}{E(\mathbf{p})}\Bigg\lbrace\int_{-\infty}^{\mu}dp^0\,S_h^n(p^0,\mathbf{p},k_F)
G^{\nu n-\bar{\nu}n}(p^0,\mathbf{p},x_N)\nonumber\\
&+&\frac{\pi^2}{2Mk_F(r)}\frac{\delta}{V}S_h^n(\mu,\mathbf{p},k_F)G^{\nu n-\bar{\nu}n}(\mu,\mathbf{p},x_N)
+\frac{\pi^2}{2k^2_F}\frac{\delta}{V}\int_{-\infty}^\mu dp^0\frac{\partial S_h^n(p^0,\mathbf{p},k)}{\partial k}\Big|_{k=k_F}G^{\nu n-
\bar{\nu}n}(p^0,\mathbf{p},x_N)\Bigg\rbrace~~\label{eq:expansion_convolution_F1_neutron}~~~
\end{eqnarray}
where $G^{\nu n-\bar{\nu}n}(p^0,\mathbf{p},x_N)$ is the same as Eq.~(\ref{eq:definition_Gnuproton}) but with the replacement of 
$p\to n$ in the structure functions.

Eqs.(\ref{eq:expansion_convolution_F1_proton}) and (\ref{eq:expansion_convolution_F1_neutron}) are summed,
keeping in mind that now the spectral functions are identical as they are evaluated at $k=k_F$, which results
\begin{eqnarray}
&2&M_A\int d^3r\int\frac{d^3p}{(2\pi)^3}\frac{M}{E(\mathbf{p})}\Bigg\lbrace\int_{-\infty}^{\mu}dp^0\,S_h(p^0,\mathbf{p},k_F)\left(G^{\nu p-\bar{\nu}p}(p^0,\mathbf{p})+G^{\nu n-\bar{\nu}n}(p^0,\mathbf{p})\right)\nonumber\\
&+&\frac{\pi^2}{2Mk_F(r)}\frac{\delta}{V}S_h(\mu,\mathbf{p},k_F)\left(G^{\nu n-\bar{\nu}n}(\mu,\mathbf{p})-G^{\nu p-\bar{\nu}p}(\mu,\mathbf{p})\right)\nonumber\\
&+&\frac{\pi^2}{2k^2_F}\frac{\delta}{V}\int_{-\infty}^\mu dp^0\frac{\partial S_h(p^0,\mathbf{p},k)}{\partial k}\Big|_{k=k_F}\left(G^{\nu n-\bar{\nu}n}(p^0,\mathbf{p})-G^{\nu p-\bar{\nu}p}(p^0,\mathbf{p})\right)\Bigg\rbrace\label{eq:sum_proton_neutron_F1}
\end{eqnarray}

The sum of the structure functions $G^{\nu n-\bar{\nu}n}(p^0,\mathbf{p})$ and $G^{\nu p-\bar{\nu}p}(p^0,\mathbf{p})$ 
in the first line of the above expression is zero in the limit of considering only light quarks 
(up- and down-quarks approximation), and  
\begin{equation}
G^{\nu n-\bar{\nu}n}(p^0,\mathbf{p})-G^{\nu p-\bar{\nu}p}(p^0,\mathbf{p})=(u_v-d_v)\left[\frac{2}{M}+\frac{4p^2_x}{M(p\cdot q)}x_N\right]
\end{equation}
where $u_v=u-\bar{u}$ and $d_v=d-\bar{d}$ are the up and down valence distributions, respectively.\\

Furthermore, in the denominator of the PW relation we have the 
difference $F_2^{\nu A}-F_2^{\bar{\nu}A}$, which is written as:
\begin{eqnarray}
 F^{A}_2&=&q^{0}W_2^{A}=q^{0}\frac{\left\langle \left( q^{2}(\vec{p}\,^{2}+2(p^{0})^{2}-p_z^{2})-
2(q^{0})^{2}\left( (p^{0})^{2}+p_z^{2}\right)+4p^{0}q^{0}p_z\sqrt{(q^{0})^{2}-q^{2}} \right)
\cdot \frac{F_2^{p}}{\nu_N} \right\rangle_{S_h^{p}} }{2M^{2}\left(q^{2}-(q^{0})^{2}\right)}\nonumber\\
&+&q^{0}\frac{\left\langle \left( q^{2}(\vec{p}\,^{2}+2(p^{0})^{2}-p_z^{2})-
2(q^{0})^{2}\left( (p^{0})^{2}+p_z^{2}\right)+4p^{0}q^{0}p_z\sqrt{(q^{0})^{2}-q^{2}} \right)
\cdot \frac{F_2^{n}}{\nu_N} \right\rangle_{S_h^{n}} }{2M^{2}\left(q^{2}-(q^{0})^{2}\right)}
\end{eqnarray}
Let us define the kinematic factor which goes in front of $F_2$ as:
\begin{equation}
 G(p^{0},\vec{p})\equiv\frac{q^{0}\left[ q^{2}(\vec{p}\,^{2}+2(p^{0})^{2}-p_z^{2})-
2(q^{0})^{2}\left( (p^{0})^{2}+p_z^{2}\right)+4p^{0}q^{0}p_z\sqrt{(q^{0})^{2}-q^{2}}\right] }
{2M\left(q^{2}-(q^{0})^{2}\right)\cdot(p\cdot q)}
\end{equation}
This expression for $F_2$ is valid for neutrino as well as for antineutrino. Therefore, when performing the subtraction, 
we get the following
\begin{equation}
 F_2^{\nu A}-F_2^{\bar{\nu}A}=\left\langle G(p^{0},\vec{p})\left(F_2^{\nu p}-F_2^{\bar{\nu}p}\right)\right\rangle_{S_h^{p}}
+ \left\langle G(p^{0},\vec{p})\left(F_2^{\nu n}-F_2^{\bar{\nu}n}\right)\right\rangle_{S_h^{n}}
\end{equation}
Defining
\begin{eqnarray}
 G_2^{\nu p-\bar{\nu}p}(p^{0},\vec{p})&\equiv&G(p^{0},\vec{p})\left(F_2^{\nu p}-F_2^{\bar{\nu}p}\right)\\
G_2^{\nu n-\bar{\nu}n}(p^{0},\vec{p})&\equiv&G(p^{0},\vec{p})\left(F_2^{\nu n}-F_2^{\bar{\nu}n}\right)
\end{eqnarray}
and using the same procedure as we did for $F_1$, one gets
\begin{eqnarray}
 \left\langle G_2^{\nu p-\bar{\nu}p}(p^{0},\vec{p})\right\rangle_{S_h^{p}}&=&
\left\langle \int_{-\infty}^{\mu}dp^{0}\, S_h^{p}(p^{0},\vec{p},k_F) 
G_2^{\nu p-\bar{\nu}p}(p^{0},\vec{p})\right\rangle\nonumber\\
&-&\left\langle \frac{\delta}{2V}\frac{\pi^{2}}{M k_F} S_h^{p}(\mu,\vec{p},k_F)
G_2^{\nu p-\bar{\nu}p}(\mu,\vec{p})\right\rangle \nonumber\\
&-&\left\langle \frac{\delta}{2V}\frac{\pi^{2}}{k_F^{2}}\int_{-\infty}^{\mu}dp^{0}\,
\frac{\partial S_h^{p}(p^{0},\vec{p},k)}{\partial k}\Big|_{k=k_F}G_2^{\nu p-\bar{\nu}p}(p^{0},\vec{p})
+\mathcal{O}\left(\frac{\delta}{V}\right)^{2}\right\rangle
\end{eqnarray}
where the symbols $\left\langle \cdots \right\rangle $ stand to indicate the 
integrals in $d^{3}p$ and $d^{3}r$. Similarly, one obtains convolution with the hole spectral function for neutrons with the 
known changes ($p\to n$ and $\delta\to-\delta$) and gets:
\begin{eqnarray}
 \left\langle G_2^{\nu n-\bar{\nu}n}(p^{0},\vec{p})\right\rangle_{S_h^{n}}&=&
\left\langle \int_{-\infty}^{\mu}dp^{0}\, S_h^{n}(p^{0},\vec{p},k_F) 
G_2^{\nu n-\bar{\nu}n}(p^{0},\vec{p})\right\rangle\nonumber\\
&+&\left\langle \frac{\delta}{2V}\frac{\pi^{2}}{M k_F} S_h^{n}(\mu,\vec{p},k_F)
G_2^{\nu n-\bar{\nu}n}(\mu,\vec{p})\right\rangle \nonumber\\
&+&\left\langle \frac{\delta}{2V}\frac{\pi^{2}}{k_F^{2}}\int_{-\infty}^{\mu}dp^{0}\,
\frac{\partial S_h^{n}(p^{0},\vec{p},k)}{\partial k}\Big|_{k=k_F}G_2^{\nu n-\bar{\nu}n}(p^{0},\vec{p})
+\mathcal{O}\left(\frac{\delta}{V}\right)^{2}\right\rangle
\end{eqnarray}
Therefore, when summing over both the contributions to obtain $F_2^{\nu A}-F_2^{\bar{\nu}A}$,
we get two kinds of contributions: the first order one and the second order one
(proportional to $\frac{\delta}{V}$).
\begin{eqnarray}
 F_2^{\nu A}-F_2^{\bar{\nu}A}&=&\left\langle G_2^{\nu p-\bar{\nu}p}(p^{0},\vec{p})\right\rangle_{S_h^{p}}+
\left\langle G_2^{\nu n-\bar{\nu}n}(p^{0},\vec{p})\right\rangle_{S_h^{n}}\nonumber\\
&=&\left\langle \int_{-\infty}^{\mu}dp^{0}\, S_h(p^{0},\vec{p},k_F)
\left( G_2^{\nu p-\bar{\nu}p}(p^{0},\vec{p})+G_2^{\nu n-\bar{\nu}n}(p^{0},\vec{p})\right) \right\rangle\nonumber\\
&+&\left\langle \frac{\delta}{2V}\frac{\pi^{2}}{M k_F} S_h(\mu,\vec{p},k_F)
\left(G_2^{\nu n-\bar{\nu}n}(p^{0},\vec{p})-G_2^{\nu p-\bar{\nu}p}(p^{0},\vec{p}) \right)  \right\rangle \nonumber\\
&+&\left\langle \frac{\delta}{2V}\frac{\pi^{2}}{k_F^{2}}\int_{-\infty}^{\mu}dp^{0}\,
\frac{\partial S_h(p^{0},\vec{p},k)}{\partial k}\Big|_{k=k_F}
\left(G_2^{\nu n-\bar{\nu}n}(p^{0},\vec{p})-G_2^{\nu p-\bar{\nu}p}(p^{0},\vec{p}) \right)
+\mathcal{O}\left(\frac{\delta}{V}\right)^{2}\right\rangle~~~~\;~ 
\end{eqnarray}
Writing the sum and the difference of $G_2$ in neutron and proton, one
obtains (in the up and down quarks approximation):
\begin{footnotesize}
\begin{eqnarray}
 G_2^{\nu p-\bar{\nu}p}(p^{0},\vec{p})+G_2^{\nu n-\bar{\nu}n}(p^{0},\vec{p})&=&
 G(p^{0},\vec{p})\left( F_2^{\nu p}-F_2^{\bar{\nu}p}+F_2^{\nu n}-F_2^{\bar{\nu}n}\right)=0\nonumber\\
 G_2^{\nu n-\bar{\nu}n}(p^{0},\vec{p})-G_2^{\nu p-\bar{\nu}p}(p^{0},\vec{p})&=&
G(p^{0},\vec{p})\left( -F_2^{\nu p}+F_2^{\bar{\nu}p}+F_2^{\nu n}-F_2^{\bar{\nu}n}\right)=
G(p^{0},\vec{p})\times 4\,x_N\left(u_v-d_v\right)~~
\end{eqnarray}
\end{footnotesize}
Therefore, in this limit, we obtain a dominant non-isoscalarity correction (because the 
isoscalar term is zero) proportional to $\frac{\delta}{V}$. This is exactly the same
effect as was obtained in the calculation of $F_1^{\nu A}-F_1^{\bar{\nu}A}$. 

For the $F_3$ structure functions for the CC case we have:
\begin{eqnarray}
 F_3^{\nu A}+F_3^{\bar{\nu} A}&=& 2A\frac{q^{0}}{q_z}\int d^{3}r\int
\frac{d^{3}p}{(2\pi)^{3}}\frac{M}{E(\vec{p})}\int_{-\infty}^{\mu_p} dp^{0}\,
S_h^{p}(p^{0},\vec{p},k_{F,p})\frac{p^{0}q_z-q^{0}p_z}{(p\cdot q)}\left( 
F_3^{\nu p}+F_3^{\bar{\nu} p}\right)\nonumber\\
&+& 2A\frac{q^{0}}{q_z}\int d^{3}r\int
\frac{d^{3}p}{(2\pi)^{3}}\frac{M}{E(\vec{p})}\int_{-\infty}^{\mu_n} dp^{0}\,
S_h^{n}(p^{0},\vec{p},k_{F,n})\frac{p^{0}q_z-q^{0}p_z}{(p\cdot q)}\left( 
F_3^{\nu n}+F_3^{\bar{\nu} n}\right)\nonumber\\
&\equiv&\left\langle G_3^{\nu p+\bar{\nu}p}(p^{0},\vec{p})\right\rangle_{S_h^{p}}
+ \left\langle G_3^{\nu n+\bar{\nu}n}(p^{0},\vec{p})\right\rangle_{S_h^{n}}
\end{eqnarray}
where in the last step we have defined:
\begin{eqnarray}
 G_3^{\nu p+\bar{\nu}p}(p^{0},\vec{p})&\equiv& 2A\,\frac{q^{0}}{q_z}\,
\frac{p^{0}q_z-q^{0}p_z}{(p\cdot q)}\left( 
F_3^{\nu p}+F_3^{\bar{\nu} p}\right)\\
G_3^{\nu n+\bar{\nu}n}(p^{0},\vec{p})&\equiv& 2A\,\frac{q^{0}}{q_z}\,
\frac{p^{0}q_z-q^{0}p_z}{(p\cdot q)}\left( 
F_3^{\nu n}+F_3^{\bar{\nu} n}\right)
\end{eqnarray}
We perform the expansion of the spectral functions around the Fermi momentum:
\begin{eqnarray}
 \left\langle G_3^{\nu p+\bar{\nu}p}(p^{0},\vec{p})\right\rangle_{S_h^{p}}&=&
\left\langle \int_{-\infty}^{\mu}dp^{0}\,S_h^{p}(p^{0},\vec{p},k_{F})
G_3^{\nu p+\bar{\nu}p}(p^{0},\vec{p})\right\rangle\nonumber\\
&-&\left\langle 
\frac{\delta}{2V}\frac{\pi^{2}}{M k_F} S_h^{p}(\mu,\vec{p},k_{F})
G_3^{\nu p+\bar{\nu}p}(\mu,\vec{p})\right\rangle\nonumber\\
&-&\left\langle  
\frac{\delta}{2V}\frac{\pi^{2}}{k_F^{2}}\int_{-\infty}^{\mu}dp^{0}\,
\frac{\partial S_h^{p}(p^{0},\vec{p},k)}{\partial k}\Big|_{k=k_F}
G_3^{\nu p+\bar{\nu}p}(p^{0},\vec{p})+\mathcal{O}\left( \frac{\delta}{V}\right)^{2}
\right\rangle \label{G3p_expansion}
\end{eqnarray}
The changes to obtain the $G_3$ convoluted structure function for neutrons are the replacements $p\to n$ and
$\delta\to-\delta$, which results
\begin{eqnarray}
 \left\langle G_3^{\nu n+\bar{\nu}n}(p^{0},\vec{p})\right\rangle_{S_h^{n}}&=&
\left\langle \int_{-\infty}^{\mu}dp^{0}\,S_h^{n}(p^{0},\vec{p},k_{F})
G_3^{\nu n+\bar{\nu}n}(p^{0},\vec{p})\right\rangle\nonumber\\
&+&\left\langle 
\frac{\delta}{2V}\frac{\pi^{2}}{M k_F} S_h^{n}(\mu,\vec{p},k_{F})
G_3^{\nu n+\bar{\nu}n}(\mu,\vec{p})\right\rangle\nonumber\\
&+&\left\langle  
\frac{\delta}{2V}\frac{\pi^{2}}{k_F^{2}}\int_{-\infty}^{\mu}dp^{0}\,
\frac{\partial S_h^{n}(p^{0},\vec{p},k)}{\partial k}\Big|_{k=k_F}
G_3^{\nu n+\bar{\nu}n}(p^{0},\vec{p})+\mathcal{O}\left( \frac{\delta}{V}\right)^{2}
\right\rangle \label{G3n_expansion}
\end{eqnarray}
Summing the above two equations for the proton and neutron, we get the following:
\begin{footnotesize}
\begin{eqnarray}
 F_3^{\nu A}+F_3^{\bar{\nu} A}&=&\left\langle \int_{-\infty}^{\mu}dp^{0}\,
S_h(p^{0},\vec{p},k_{F})\left(G_3^{\nu p+\bar{\nu}p}(p^{0},\vec{p})+
G_3^{\nu n+\bar{\nu}n}(p^{0},\vec{p})\right)\right\rangle\nonumber\\
&+&\left\langle 
\frac{\delta}{2V}\frac{\pi^{2}}{M k_F} S_h(\mu,\vec{p},k_{F})\left(
G_3^{\nu n+\bar{\nu}n}(\mu,\vec{p})-G_3^{\nu p+\bar{\nu}p}(\mu,\vec{p})
\right)\right\rangle\nonumber\\
&+&\left\langle  
\frac{\delta}{2V}\frac{\pi^{2}}{k_F^{2}}\int_{-\infty}^{\mu}dp^{0}\,
\frac{\partial S_h(p^{0},\vec{p},k)}{\partial k}\Big|_{k=k_F}\left(
G_3^{\nu n+\bar{\nu}n}(p^{0},\vec{p})-G_3^{\nu p+\bar{\nu}p}(p^{0},\vec{p})
\right)+\mathcal{O}\left( \frac{\delta}{V}\right)^{2}
\right\rangle~~\label{sum_proton_neutron_F3}
\end{eqnarray}
\end{footnotesize}
If we write explicitly the combinations of $G_3$ functions in terms of
PDFs considering only light quarks (up and down), we get:
\begin{equation}
 G_3^{\nu p+\bar{\nu}p}(p^{0},\vec{p})+G_3^{\nu n+\bar{\nu}n}(p^{0},\vec{p})=
4G_3(p^{0},\vec{p})\left(u_v+d_v\right)
\end{equation}
where 
\begin{equation}\label{G3_function}
 G_3(p^{0},\vec{p})\equiv 2A\,\frac{q^{0}}{q_z}\,
\frac{p^{0}q_z-q^{0}p_z}{(p\cdot q)}
\end{equation}
And for the difference we get exactly zero i.e.
\begin{equation}
 G_3^{\nu n+\bar{\nu}n}(p^{0},\vec{p})-G_3^{\nu p+\bar{\nu}p}(p^{0},\vec{p})=0
\end{equation}
Therefore, for $F_3$, the dominant contribution is the isoscalar one (the one 
which is not proportional to $\frac{\delta}{V}$) and the non-isoscalar contribution
(proportional to $\frac{\delta}{V}$) is zero at first order ($\mathcal{O}
\left(\frac{\delta}{V}\right)$). Of course there will be corrections at higher
orders, but we are retaining only the first order non-isoscalar corrections.\\

For neutral currents, in the numerator of the PW relation, we would only
have the structure function $F_3^{0A}$, which can be written as:
\begin{eqnarray}
 F_3^{0A}(x_A)&=& 2A\,\frac{q^{0}}{q_z}\int d^{3}r\int \frac{d^{3}p}{(2\pi)^{3}}
\frac{M}{E(\vec{p})}\int_{-\infty}^{\mu_p}dp^{0}\,S_h^{p}(p^{0},\vec{p},k_{F,p})
\frac{p^{0} q_z-p_z q^{0}}{p\cdot q}F_3^{0p}(x_N)\nonumber\\
&+& 2A\,\frac{q^{0}}{q_z}\int d^{3}r\int \frac{d^{3}p}{(2\pi)^{3}}
\frac{M}{E(\vec{p})}\int_{-\infty}^{\mu_n}dp^{0}\,S_h^{n}(p^{0},\vec{p},k_{F,n})
\frac{p^{0} q_z-p_z q^{0}}{p\cdot q}F_3^{0n}(x_N)\\
&\equiv&\left\langle G_3^{0p}(p^{0},\vec{p})\right\rangle_{S_h^{p}}+
 \left\langle G_3^{0n}(p^{0},\vec{p})\right\rangle_{S_h^{n}}
\end{eqnarray}
where for the neutral current case, we define
\begin{eqnarray}
G_3^{0p}(p^{0},\vec{p})\equiv G_3(p^{0},\vec{p})F_3^{0p}(x_N)\\
G_3^{0n}(p^{0},\vec{p})\equiv G_3(p^{0},\vec{p})F_3^{0n}(x_N).
\end{eqnarray}
$G_3(p^{0},\vec{p})$ is defined in Eq.~(\ref{G3_function}).\\

The expression for $F_3^{0A}$ around the isoscalar condition is:
\begin{eqnarray}
 F_3^{0A}&=&\left\langle \int_{-\infty}^{\mu}dp^{0}\,
S_h(p^{0},\vec{p},k_{F})\left(G_3^{0p}(p^{0},\vec{p})+
G_3^{0n}(p^{0},\vec{p})\right)\right\rangle\nonumber\\
&+&\left\langle 
\frac{\delta}{2V}\frac{\pi^{2}}{M k_F} S_h(\mu,\vec{p},k_{F})\left(
G_3^{0n}(\mu,\vec{p})-G_3^{0p}(\mu,\vec{p})
\right)\right\rangle\nonumber\\
&+&\left\langle  
\frac{\delta}{2V}\frac{\pi^{2}}{k_F^{2}}\int_{-\infty}^{\mu}dp^{0}\,
\frac{\partial S_h(p^{0},\vec{p},k)}{\partial k}\Big|_{k=k_F}\left(
G_3^{0n}(p^{0},\vec{p})-G_3^{0p}(p^{0},\vec{p})
\right)+\mathcal{O}\left( \frac{\delta}{V}\right)^{2}
\right\rangle
\end{eqnarray}
For the neutral current in the parton model
\begin{equation}
 F_3^{0p}=2\left[\left( \epsilon_L^{2}(u)-\epsilon_R^{2}(u)\right)
\left( u-\bar{u}\right)+\left( \epsilon_L^{2}(d)-\epsilon_R^{2}(d)\right)
\left( d-\bar{d}\right)   \right]\equiv 2\left[g^{2}_{-}(u)
\left( u-\bar{u}\right)+ g^{2}_{-}(d)
\left( d-\bar{d}\right)\right]  
\end{equation}
where
\begin{eqnarray}
 g^{2}_{-}(u)&\equiv&\epsilon_L^{2}(u)-\epsilon_R^{2}(u)=\frac{1}{4}-
\frac{2}{3}\sin^{2}\theta_W\\
g^{2}_{-}(d)&\equiv&\epsilon_L^{2}(d)-\epsilon_R^{2}(d)=\frac{1}{4}-
\frac{1}{3}\sin^{2}\theta_W
\end{eqnarray}
and invoking isospin symmetry, we also have:
\begin{equation}
 F_3^{0n}=2\left[g^{2}_{-}(u)
\left( d-\bar{d}\right)+ g^{2}_{-}(d)
\left( u-\bar{u}\right)\right]  
\end{equation}
With these, we obtain the expressions for the sum and
difference of neutral currents $G_3$ functions for protons and neutrons i.e.
\begin{equation}
 G_3^{0p}(p^{0},\vec{p})+
G_3^{0n}(p^{0},\vec{p})=2\, G_3(p^{0},\vec{p})\left( g^{2}_{-}(u)+
g^{2}_{-}(d)\right)\left( u_v+d_v\right)  
\end{equation}
\begin{equation}
 G_3^{0n}(p^{0},\vec{p})-G_3^{0p}(p^{0},\vec{p})=2\, G_3(p^{0},\vec{p})
\left(g^{2}_{-}(d)- g^{2}_{-}(u)\right) \left(u_v-d_v \right) 
\end{equation}

\subsection{Nuclear corrections to PW ratio}\label{subsec:PW}
Using the expressions obtained in the previous section, we may write the ratio of the differential scattering cross sections as
\begin{eqnarray}
 R^-&=&\frac{\frac{d\sigma_{NC}^{\nu A}}{dx\,dy}-\frac{d\sigma_{NC}^{\bar{\nu}A}}{dx\,dy}}{\frac{d\sigma_{CC}^
 {\nu A}}{dx\,dy}-\frac{d\sigma_{CC}^{\bar{\nu}A}}{dx\,dy}} \nonumber \\
 &=&\frac{2xy\left(1-\frac{y}{2} \right)\left[2\left(g^{2}_{-}(u)+
 g^{2}_{-}(d)\right)\left\langle \int_{-\infty}^{\mu}dp^{0}\,S_h\,G_3^{\prime}
\left(u_v(x_N)+d_v(x_N) \right)  \right\rangle +\mathcal{O}(\epsilon_3^{0A})
 \right]  }{y^{2}x\,\mathcal{O}(\epsilon_1^{\nu A-\bar{\nu}A})+
\left\lbrace 1-y-\frac{Mxy}{2E_\nu}\right\rbrace 
\mathcal{O}(\epsilon_2^{\nu A-\bar{\nu}A})+xy\left(1-\frac{y}{2} \right)
\left\langle 4\int_{-\infty}^{\mu}dp^{0}\,S_h\,G_3^{\prime}
(u_v+d_v) \right\rangle }\nonumber\\
&=&\frac{4xy\left(1-\frac{y}{2} \right)\left(g^{2}_{-}(u)+
 g^{2}_{-}(d)\right)\left\langle \int_{-\infty}^{\mu}dp^{0}\,S_h\,G_3^{\prime}
\left(u_v(x_N)+d_v(x_N) \right) \right\rangle+\mathcal{O}^{\prime}(\epsilon_3^{0A})}
{4xy\left(1-\frac{y}{2} \right)\left\langle \int_{-\infty}^{\mu}dp^{0}\,S_h\,G_3^{\prime}
\left(u_v(x_N)+d_v(x_N) \right) \right\rangle+\mathcal{O}^{\prime}
(\epsilon^{\nu A-\bar{\nu}A})}~~~\label{eq:PW_ratio2}
\end{eqnarray}
where
\begin{eqnarray}
G_3^\prime&=&\frac{G_3(p^0,\vec{p})}{A}\\
 \mathcal{O}^{\prime}(\epsilon_3^{0A})&=&2xy\left(1-\frac{y}{2} \right)
\mathcal{O}(\epsilon_3^{0A})\\
\mathcal{O}^{\prime}
(\epsilon^{\nu A-\bar{\nu}A})&=&y^{2}x\;\mathcal{O}(\epsilon_1^{\nu A-\bar{\nu}A})+
\left\lbrace 1-y-\frac{Mxy}{2E_\nu}\right\rbrace 
\mathcal{O}(\epsilon_2^{\nu A-\bar{\nu}A})
\end{eqnarray}
Here $\epsilon$ indicates that the contribution is of the order $\mathcal{O}\left(\frac{\delta}{V}\right)$.

Expanding the denominator of Eq.~(\ref{eq:PW_ratio2}) using Taylor series and keeping only the first order term in $\frac{\delta}{V}$, we may
write the above equation as:
\begin{eqnarray} \label{correctioninPW}
 R^{-}&\simeq&\left[4xy\left(1-\frac{y}{2} \right)\left(g^{2}_{-}(u)+
 g^{2}_{-}(d)\right)\left\langle \int_{-\infty}^{\mu}dp^{0}\,S_h\,G_3^{\prime}
\left(u_v+d_v \right) \right\rangle+\mathcal{O}^{\prime}(\epsilon_3^{0A})\right]
\nonumber\\ 
&\times&\left[\frac{1}{4xy\left(1-\frac{y}{2} \right)\left\langle 
\int_{-\infty}^{\mu}dp^{0}\,S_h\,G_3^{\prime}
\left(u_v+d_v \right) \right\rangle}-\frac{\mathcal{O}^{\prime}
(\epsilon^{\nu A-\bar{\nu}A})}{\left[4xy\left(1-\frac{y}{2} \right)\left\langle 
\int_{-\infty}^{\mu}dp^{0}\,S_h\,G_3^{\prime}
\left(u_v+d_v \right) \right\rangle \right]^{2}} \right]\nonumber\\
&\simeq&g^{2}_{-}(u)+g^{2}_{-}(d)+\frac{\mathcal{O}^{\prime}(\epsilon_3^{0A})}
{4xy\left(1-\frac{y}{2} \right)\left\langle 
\int_{-\infty}^{\mu}dp^{0}\,S_h\,G_3^{\prime}
\left(u_v+d_v \right) \right\rangle}\nonumber\\
&-&\frac{\left( g^{2}_{-}(u)+g^{2}_{-}(d)
\right)\mathcal{O}^{\prime}
(\epsilon^{\nu A-\bar{\nu}A}) }{4xy
\left(1-\frac{y}{2} \right)\left\langle 
\int_{-\infty}^{\mu}dp^{0}\,S_h\,G_3^{\prime}
\left(u_v+d_v \right) \right\rangle}\nonumber\\
&=&\frac{1}{2}-\sin^{2}\theta_W + \delta R^{-}
\end{eqnarray}
where $\delta R^{-}$ is the correction due to the non-isoscalarity of
the target and is written as
\begin{equation}\label{deltaRminus}
\delta R^{-}=\delta R^{-}_1~+~\delta R^{-}_2.\;\;
\end{equation}
The first correction $\delta R^{-}_1$ is given by:
\begin{eqnarray}\label{firstcorrection}
 \delta R^{-}_1&=&\frac{\mathcal{O}^{\prime}(\epsilon_3^{0A})}
{4xy\left(1-\frac{y}{2} \right)\left\langle 
\int_{-\infty}^{\mu}dp^{0}\,S_h\,G_3^{\prime}
\left(u_v+d_v \right) \right\rangle}=\frac{\mathcal{O}(\epsilon_3^{0A})}
{2\left\langle 
\int_{-\infty}^{\mu}dp^{0}\,S_h\,G_3^{\prime}
\left(u_v+d_v \right) \right\rangle}\nonumber\\
&=&\frac{\left\langle\frac{\delta}{2V}\frac{\pi^{2}}{M k_F}
S_h(\mu,\vec{p},k_F)(G_3^{\prime0n}(\mu,\vec{p})-G_3^{\prime0p}(\mu,\vec{p}))
+ \frac{\delta}{2V}\frac{\pi^{2}}{k^{2}_F}\int_{-\infty}^{\mu}dp^{0}\,
\frac{\partial S_h}{\partial k}\Big|_{k=k_F}
(G_3^{\prime0n}(p^{0},\vec{p})-G_3^{\prime0p}(p^{0},\vec{p}))
\right\rangle }{2\left\langle 
\int_{-\infty}^{\mu}dp^{0}\,S_h\,G_3^{\prime}
\left(u_v+d_v \right) \right\rangle}\nonumber\\
&=&\frac{(g^{2}_{-}(d)-g^{2}_{-}(u))\left\langle\frac{\delta}{2V}
\frac{\pi^{2}}{M k_F}(S_h\,G_3^{\prime}(u_v-d_v))\Big|_{p^{0}=\mu}
+ \frac{\delta}{2V}\frac{\pi^{2}}{k^{2}_F}\int_{-\infty}^{\mu}dp^{0}\,
\frac{\partial S_h}{\partial k}\Big|_{k=k_F}G_3^{\prime}(u_v-d_v)
\right\rangle }{\left\langle 
\int_{-\infty}^{\mu}dp^{0}\,S_h\,G_3^{\prime}
\left(u_v+d_v \right) \right\rangle}
\end{eqnarray}
and the second correction $\delta R^{-}_2$  is given by:
\begin{eqnarray}\label{secondcorrection}
 \delta R^{-}_2&=&-\left(g^{2}_{-}(u)+g^{2}_{-}(d) \right)\,
\frac{\mathcal{O}^{\prime}(\epsilon^{\nu A-\bar{\nu}A})}
{4xy\left( 1-\frac{y}{2}\right)\left\langle \int_{-\infty}^{\mu}dp^{0}\,
S_h G_3^{\prime}(u_v+d_v) \right\rangle } \nonumber\\
&=&-\left(g^{2}_{-}(u)+g^{2}_{-}(d) \right)\,
\frac{y^{2}x\,\mathcal{O}(\epsilon_1^{\nu A-\bar{\nu}A})+
\left\lbrace 1-y-\frac{Mxy}{2E_\nu}\right\rbrace 
\mathcal{O}(\epsilon_2^{\nu A-\bar{\nu}A})}{4xy\left( 1-\frac{y}{2}\right)
\left\langle \int_{-\infty}^{\mu}dp^{0}\,
S_h G_3^{\prime}(u_v+d_v) \right\rangle }=\delta R^{-}_{2,1}+\delta R^{-}_{2,2}\;\;~~~
\end{eqnarray}
where
\begin{eqnarray}
 \delta R^{-}_{2,1}&=&-\left(g^{2}_{-}(u)+g^{2}_{-}(d) \right)\,
\frac{y^{2}x\,\mathcal{O}(\epsilon_1^{\nu A-\bar{\nu}A})}{4xy
\left( 1-\frac{y}{2}\right)
\left\langle \int_{-\infty}^{\mu}dp^{0}\,
S_h G_3^{\prime}(u_v+d_v) \right\rangle }\label{deltaR21}\\
\delta R^{-}_{2,2}&=&-\left(g^{2}_{-}(u)+g^{2}_{-}(d) \right)\,
\frac{1-y-\frac{Mxy}{2E_\nu}}{4xy\left( 1-\frac{y}{2}\right)}\,
\frac{\mathcal{O}(\epsilon_2^{\nu A-\bar{\nu}A})}
{\left\langle \int_{-\infty}^{\mu}dp^{0}\,
S_h G_3^{\prime}(u_v+d_v) \right\rangle }\label{deltaR22}
\end{eqnarray}
Making the substitution of $\mathcal{O}(\epsilon_1^{\nu A-\bar{\nu}A})$
in Eq.~(\ref{deltaR21}), we get:
\begin{eqnarray}
 \delta R^{-}_{2,1}&=&-\left(g^{2}_{-}(u)+g^{2}_{-}(d) \right)\,
\frac{y\, 2M\left\langle\frac{\delta}{2V}\frac{\pi^{2}}{Mk_F}
S_h(\mu,\vec{p},k_F)\left(G_1^{\nu n-\bar{\nu}n}-G_1^{\nu p-\bar{\nu}p}
 \right)_{p^{0}=\mu}\right\rangle}{4\left( 1-\frac{y}{2}\right)
\left\langle \int_{-\infty}^{\mu}dp^{0}\,
S_h G_3^{\prime}(u_v+d_v) \right\rangle}\nonumber\\
 &-&\left(g^{2}_{-}(u)+g^{2}_{-}(d) \right)\,
\frac{y\,2M\left\langle\frac{\delta}{2V}
\frac{\pi^{2}}{k_F^{2}}\int_{-\infty}^{\mu}dp^{0}\,
\frac{\partial S_h}{\partial k}\Big|_{k=k_F}\left(G_1^{\nu n-\bar{\nu}n}-
G_1^{\nu p-\bar{\nu}p} \right)\right\rangle }{4\left( 1-\frac{y}{2}\right)
\left\langle \int_{-\infty}^{\mu}dp^{0}\,
S_h G_3^{\prime}(u_v+d_v) \right\rangle }
\end{eqnarray}
In the parton model, we have:
\begin{equation}
 G_1^{\nu n-\bar{\nu}n}-G_1^{\nu p-\bar{\nu}p}=\frac{1}{M}(u_v-d_v)
\left(2+\frac{4p_x^{2}\,x_N}{(p\cdot q)} \right)=\frac{1}{M}(u_v-d_v)\,
G_1(p^{0},\vec{p})
\end{equation}
With this, we can write:
\begin{eqnarray}
 \delta R^{-}_{2,1}&=&-\left(g^{2}_{-}(u)+g^{2}_{-}(d) \right)\,
\frac{y}{2-y}\frac{\left\langle\frac{\delta}{2V}\left[\frac{\pi^{2}}{Mk_F}
\left(S_h G_1(u_v-d_v) \right)\Big|_{p^{0}=\mu}+\frac{\pi^{2}}{k_F^{2}}
\int_{-\infty}^{\mu}dp^{0}\,
\frac{\partial S_h}{\partial k}\Big|_{k=k_F} G_1\,(u_v-d_v)\right]  \right\rangle}
{\left\langle \int_{-\infty}^{\mu}dp^{0}\,
S_h G_3^{\prime}(u_v+d_v) \right\rangle }\nonumber\\
\end{eqnarray}
Similarly using $G_2^{\nu n-\bar{\nu}n}-G_2^{\nu p-\bar{\nu}p}=G_2(p^{0},\vec{p})\cdot 4
x_N(u_v-d_v)$, we may write

\begin{eqnarray}
\delta R^{-}_{2,2}&=&-\left(g^{2}_{-}(u)+g^{2}_{-}(d) \right)\,
\frac{1-y-\frac{Mxy}{2E_\nu}}{xy\left(1-\frac{y}{2}\right)} \times \nonumber \\ 
&&\frac{\left\langle\frac{\delta}{2V}\left[\frac{\pi^{2}}{Mk_F}
\left(S_h G_2\,x_N(u_v-d_v) \right)\Big|_{p^{0}=\mu}+
\frac{\pi^{2}}{k_F^{2}}\int_{-\infty}^{\mu}dp^{0}\,
\frac{\partial S_h}{\partial k}\Big|_{k=k_F} G_2\,x_N (u_v-d_v)\right]
\right\rangle}
{\left\langle \int_{-\infty}^{\mu}dp^{0}\,
S_h G_3^{\prime}(u_v+d_v) \right\rangle }\;\;~~
\end{eqnarray}

We must point out that when $p^{0}=\mu$, $S_h(\mu,\vec{p},k_F)$ is zero 
\cite{Marco:1995vb,FernandezdeCordoba:1991wf} and the imaginary part of the self-energy of 
the nucleon is also zero, and this has been used in the definition (numerator) of the 
hole spectral function while performing the numerical calculations. 

\begin{figure}[h]
\includegraphics[width=14cm,height=10cm]{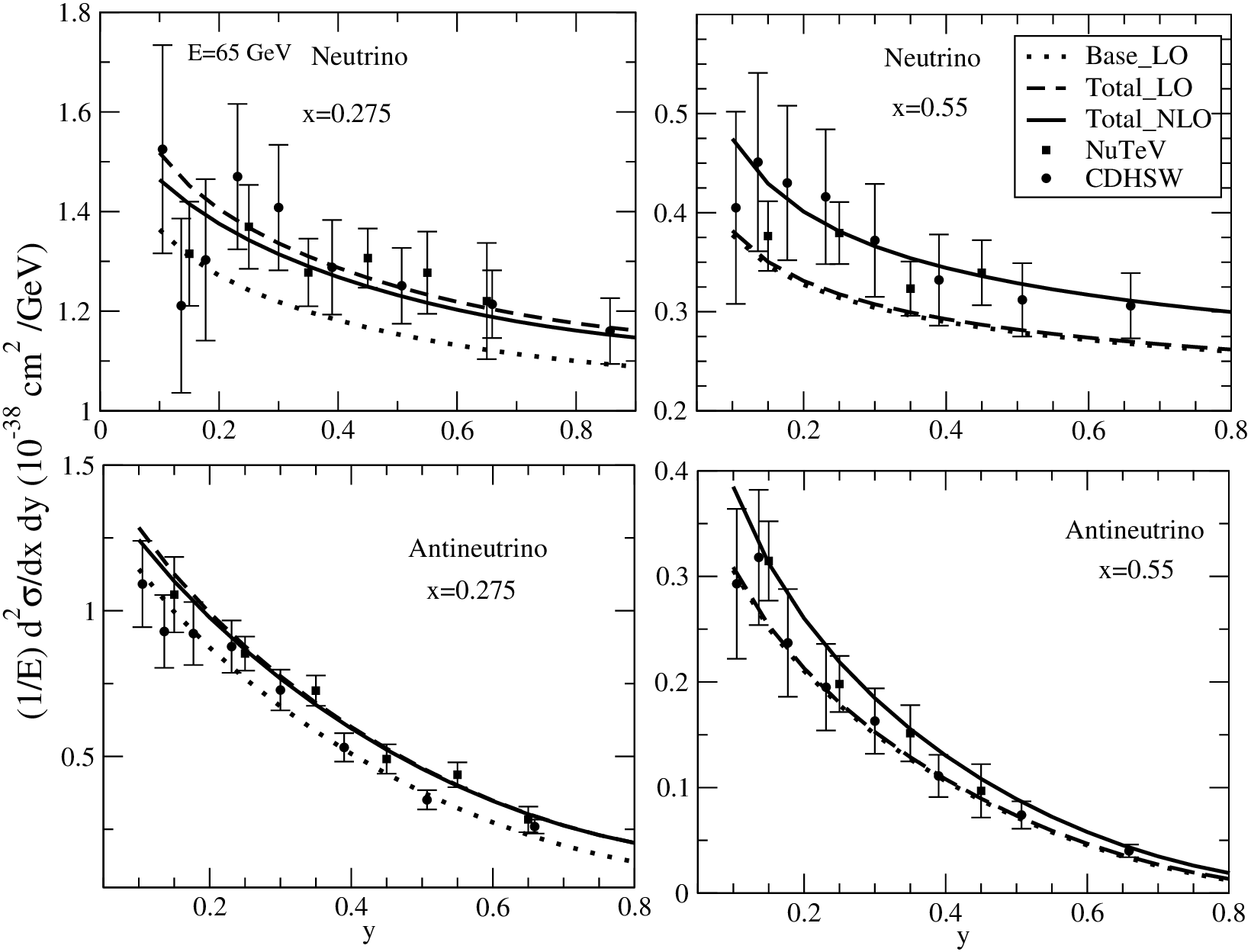}
\caption{$\frac{1}{E}\frac{d^2\sigma}{dxdy}$ vs y at different x for charged current $\nu_\mu(\bar\nu_\mu) $($E_{\nu_\mu}=65$ GeV) induced reaction in $^{56}$Fe. Dotted line is the base results and for numerical calculations CTEQ~\cite{cteq} PDFs at LO have been used. 
Dashed line is the full model at LO. Solid line is full calculation at NLO. The experimental points are from CDHSW~\cite{Berge}(solid circle) and NuTeV~\cite{Tzanov}(solid square) experiments.}
\label{d2sigma_cc_nu_nubar}
\end{figure}

\begin{figure}[h]
\includegraphics[width=14cm,height=10cm]{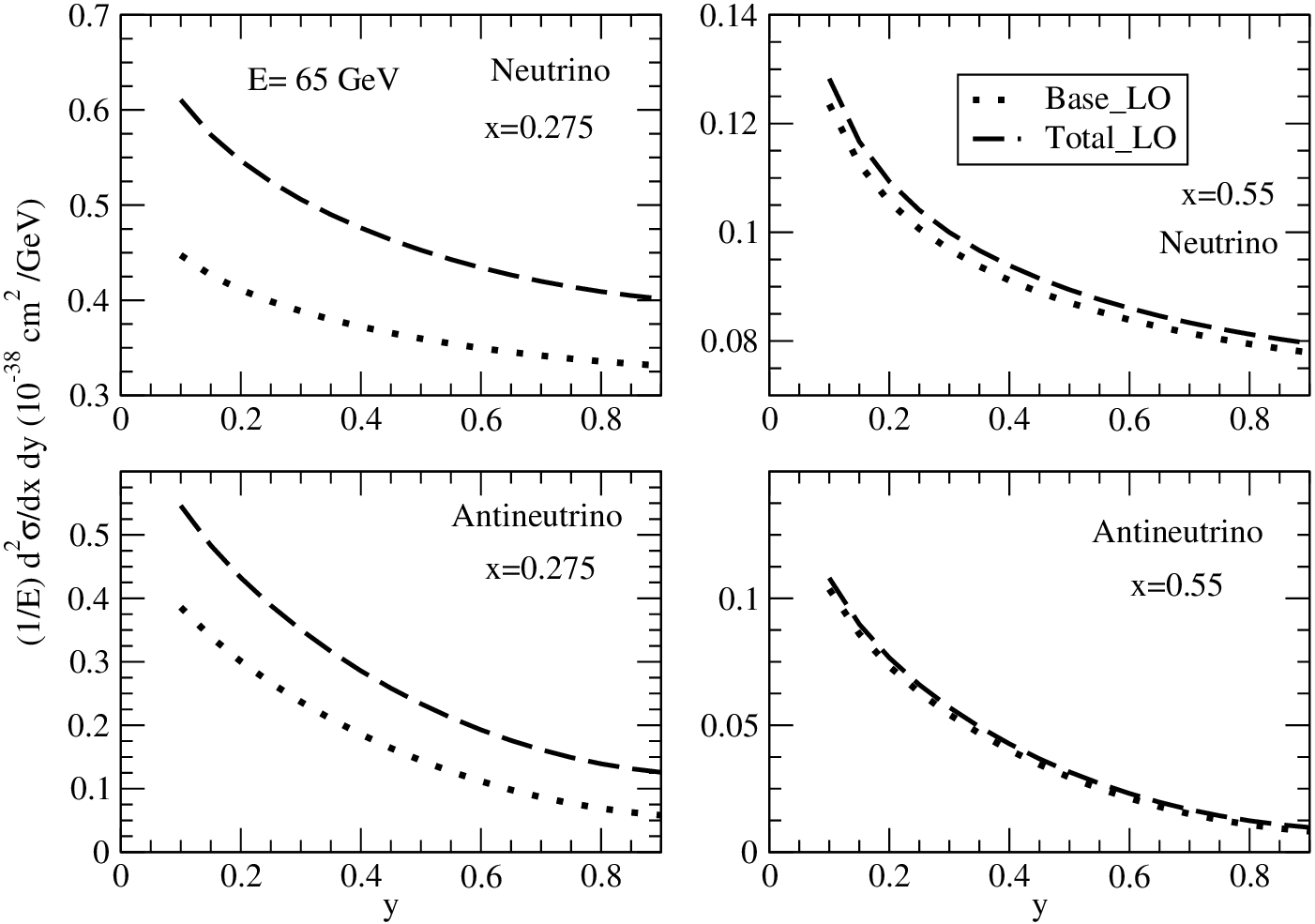}
\caption{$\frac{1}{E}\frac{d^2\sigma}{dxdy}$ vs y at different x for neutral current $\nu_\mu(\bar\nu_\mu) $($E_{\nu_\mu}=65$ GeV) induced reaction in $^{56}$Fe. Lines have the same meaning as in fig.\ref{d2sigma_cc_nu_nubar}.}
\label{d2sigma_nc_nu_nubar}
\end{figure}

\begin{figure}[b]
\includegraphics[width=14cm,height=10cm]{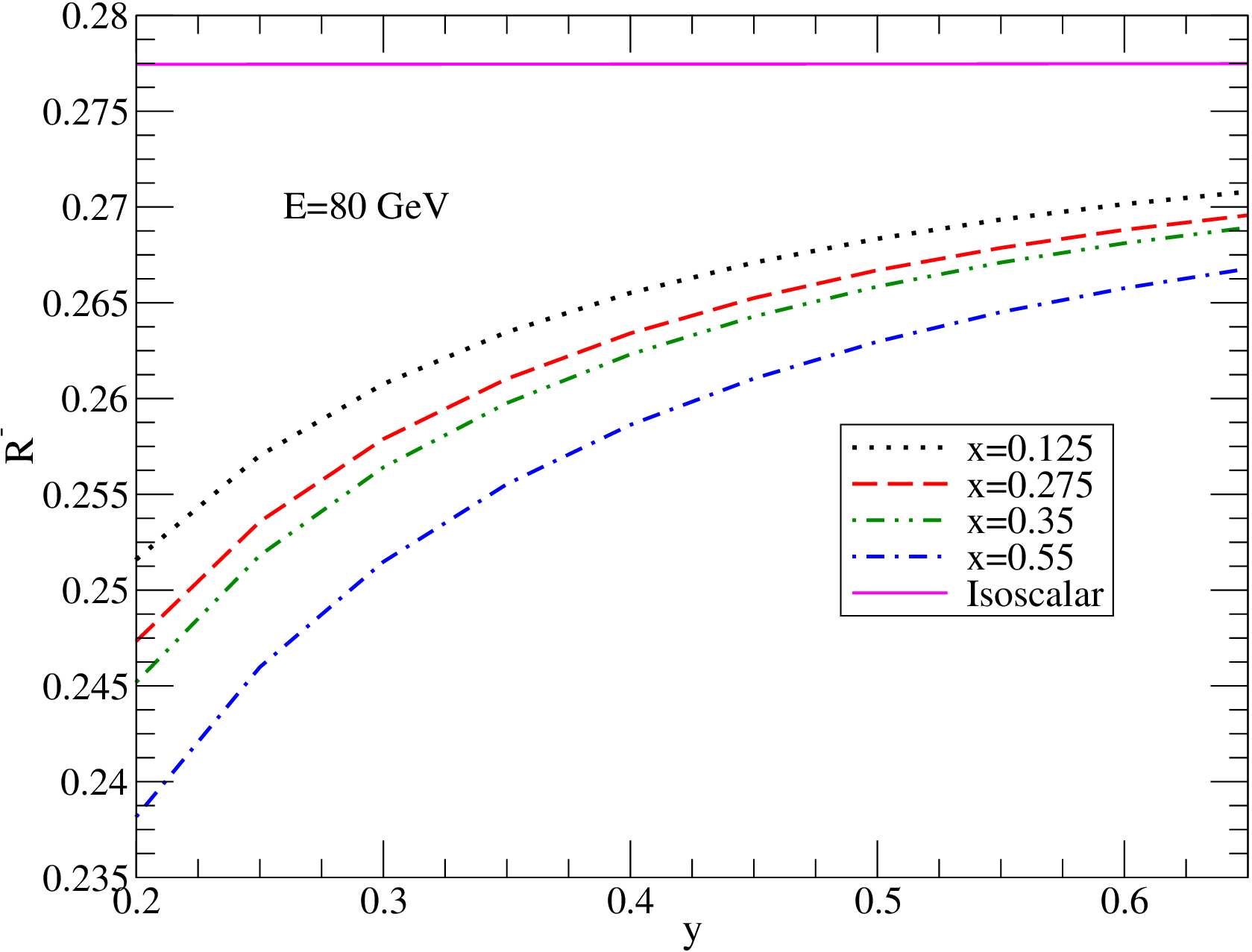}
\caption{Paschos and Wolfenstein ratio $R^-=\frac{\frac{d^2\sigma(\nu_\mu~N \rightarrow \nu_\mu~X)}{dxdy}~-~\frac{d^2\sigma(\bar\nu_\mu~N \rightarrow \bar\nu_\mu~X)}{dxdy}}
{\frac{d^2\sigma(\nu_\mu~N \rightarrow \mu^-~X)}{dxdy}~-~\frac{d^2\sigma(\bar\nu_\mu~N \rightarrow \mu^+~X)}{dxdy}}$ in $^{56}$Fe for non-isoscalar case.  $R^-$ is calculated for 
(anti)neutrino energy(E) of 80 GeV and at different Bjorken x. The solid line is the result when one treats $^{56}$Fe to be an isoscalar target.}
\label{rminus}
\end{figure}

\begin{figure}[h]
\includegraphics[width=14cm,height=10cm]{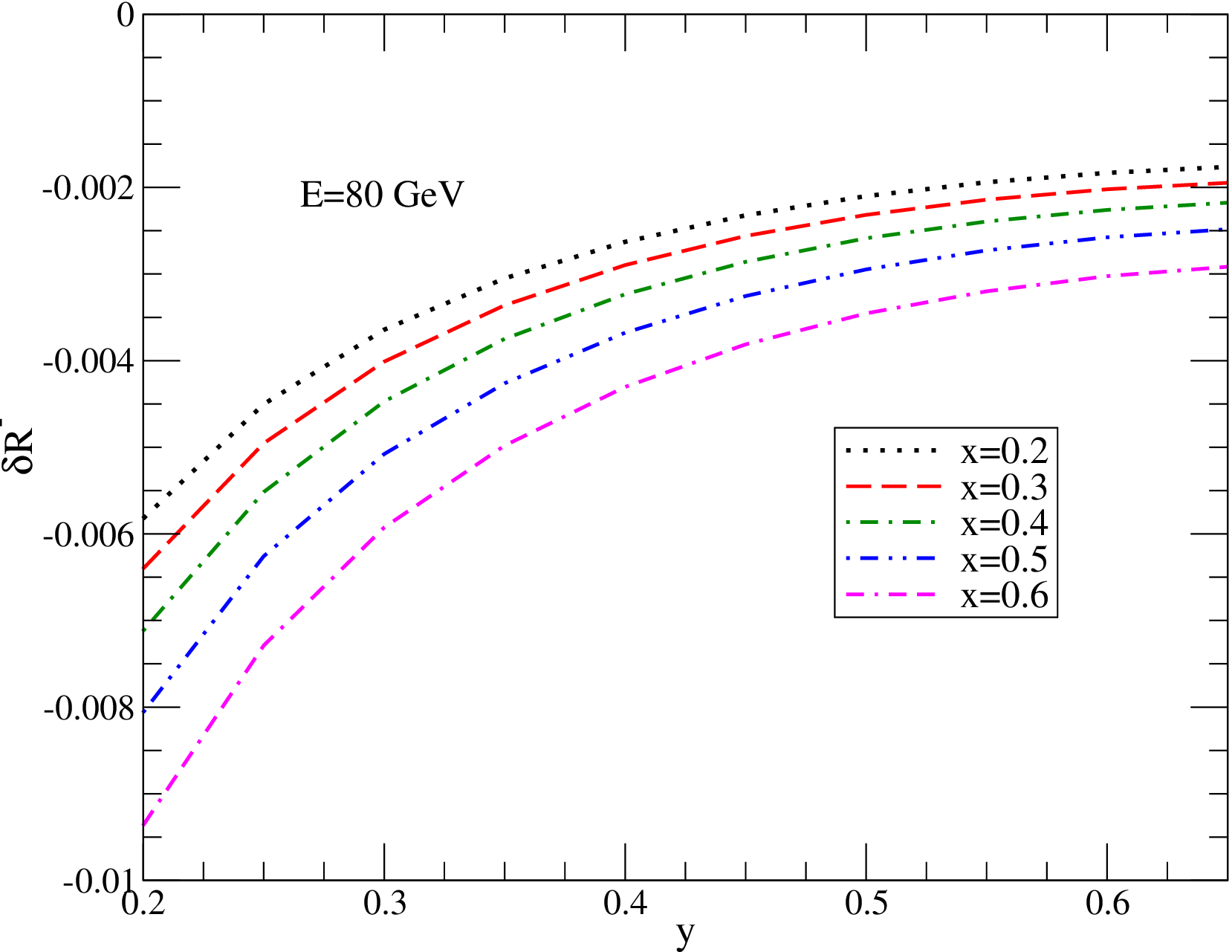}
\caption{Nonisoscalar correction($\delta R^-$) vs y at different values of x for (anti)neutrino energy(E) of 80 GeV.}
\label{deltaR}
\end{figure}

\begin{figure}[h]
\includegraphics[width=14cm,height=10cm]{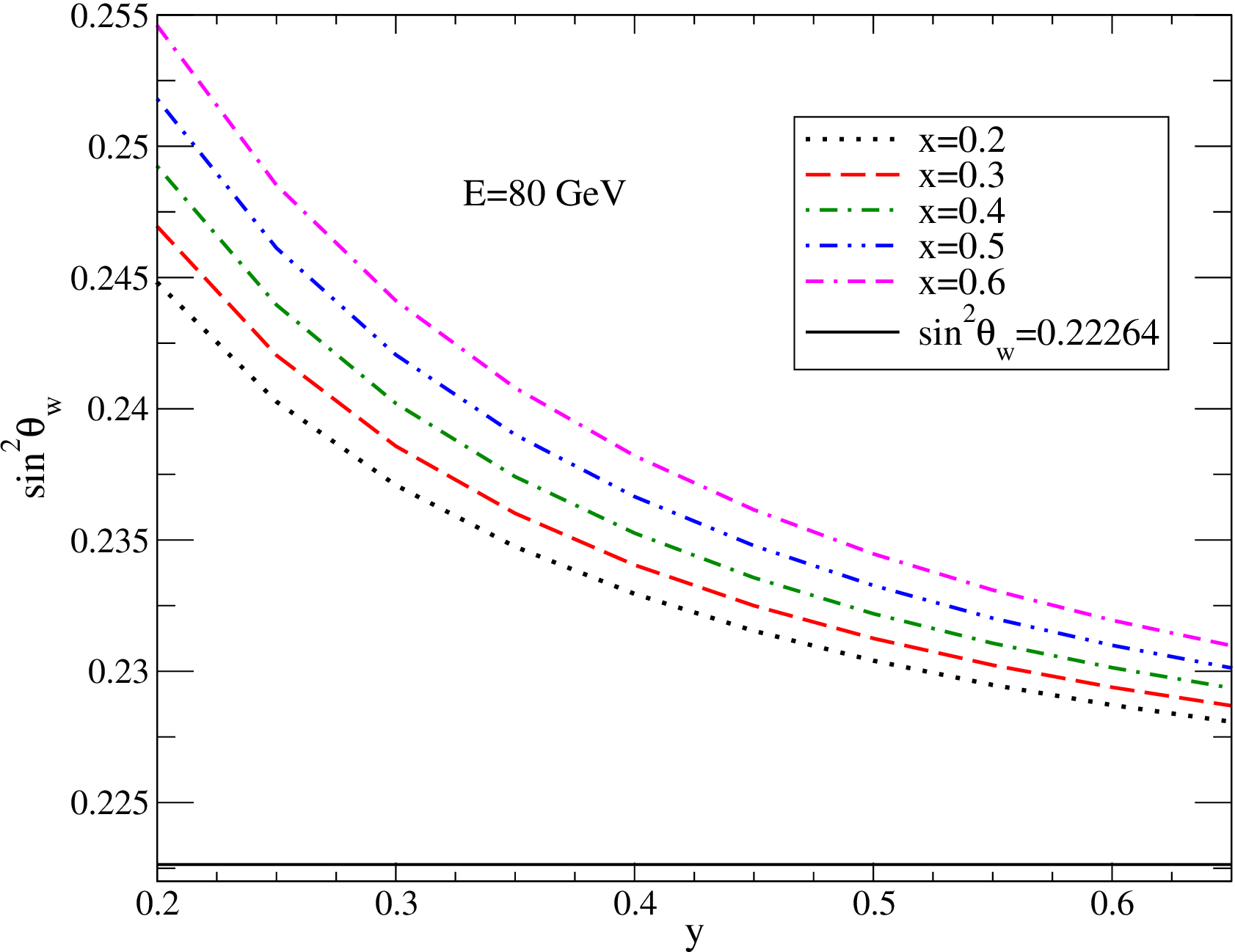}
\caption{$sin^{2}\theta_w$ vs y at different values of x in $^{56}$Fe treating it to be non-isoscalar nuclear target for (anti)neutrino energy(E) of 80 GeV.}
\label{sintheta}
\end{figure}

\begin{figure}[h]
\includegraphics[width=14cm,height=10cm]{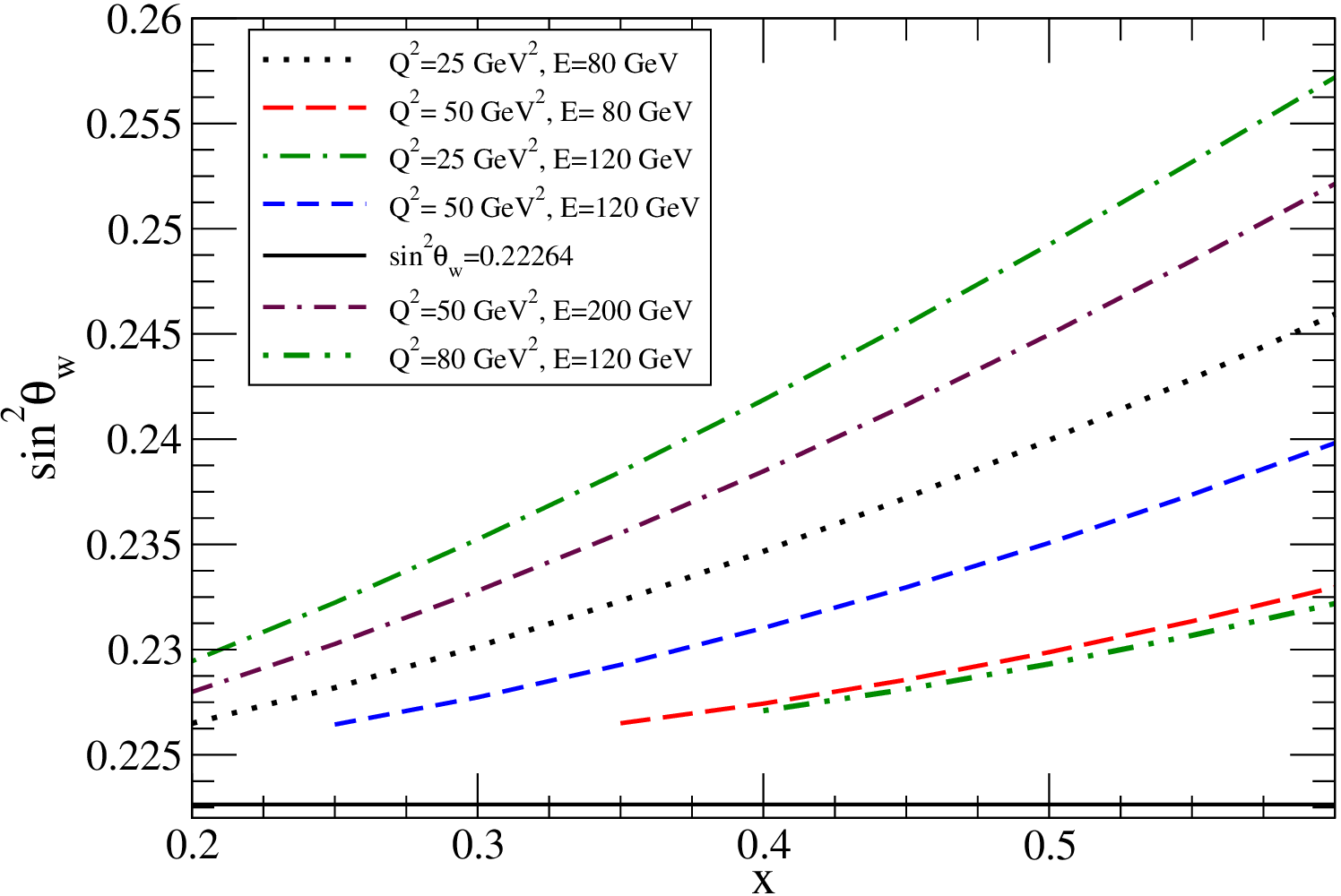}
\caption{$sin^{2}\theta_w$ vs x in $^{56}$Fe treating it to be non-isoscalar nuclear target. The results are shown at different values of $Q^2$ and (anti)neutrino energy E.}
\label{sintheta_EQ}
\end{figure}

\section{Results and Discussion}
\label{sec:result}
In this section we present and discuss the results of our numerical calculations. For performing the numerical calculations we have used the expressions for the weak structure functions 
$F^A_2$ and $F^A_3$ for an isoscalar target~\cite{Sajjad2}:  
\begin{eqnarray}\label{f2Anuclei}
F^A_2(x_A,Q^2)&=&4\int d^3r\int \frac{d^3p}{(2\pi)^3}\frac{M}{E({\bf p})}\int_{-\infty}^\mu dp^0\;
S_h(p^0,\mathbf{p},\rho(\mathbf{r})) \frac{x}{x_N}
\left( 1+\frac{2x_N p_x^2}{M\nu_N} \right)  F_2^N(x_N,Q^2)\;\;~~~~
\end{eqnarray}
\begin{eqnarray}\label{f3Anuclei}
F_3^A(x_A,Q^2)&=&4\int d^3r \int \frac{d^3p}{(2\pi)^3} \frac{M}{E({\bf p})}\int_{-\infty}^{\mu} dp^0
S_h(p^0,\mathbf{p},\rho(\mathbf{r})) \frac{p^0\gamma-p_z}{(p^0-p_z\gamma)\gamma} F_3^N(x_N,Q^2)
\end{eqnarray}
and for a nonisoscalar target, the expressions for $F^A_2(x)$ and $F^A_3(x)$ are given by~\cite{Sajjad3}:
\begin{eqnarray}\label{f2Anuclei}
F^A_2(x_A,Q^2)&=& 2\int d^3r\int\frac{d^3p}{(2\pi)^3}\frac{M}{E(\mathbf{p})}\left[ \int^{\mu_p}_{-\infty}dp^0\; S^{proton}_{h}(p^0,\mathbf{p},k_{F,p}) F_2^{proton}(x_N,Q^2) \right. \nonumber   \\ 
 &+& \left. \int^{\mu_n}_{-\infty}dp^0\; S^{neutron}_{h}(p^0,\mathbf{p},k_{F,n}) F_2^{neutron}(x_N,Q^2) \right] \frac{x}{x_N} \left(1+\frac{2x_N p_x^2}{M\nu_N}\right)  
\end{eqnarray}
\begin{eqnarray}\label{f3Anuclei}
F_3^A(x_A,Q^2)&=& 2\int d^3r\int\frac{d^3p}{(2\pi)^3}\frac{M}{E(\mathbf{p})}\left[\int^{\mu_p}_{-\infty}dp^0\; S^{proton}_{h}(p^0,\mathbf{p},k_{F,p}) F_3^{proton}(x_N,Q^2) \right. \nonumber \\
 &+& \left. \int^{\mu_n}_{-\infty}dp^0\; S^{neutron}_{h}(p^0,\mathbf{p},k_{F,n}) F_3^{neutron}(x_N,Q^2) \right] \frac{p^0\gamma-p_z}{(p^0-p_z\gamma)\gamma} 
\end{eqnarray}
where 
\begin{equation}	\label{gamma}
\gamma=\frac{q_z}{q^0}=
\left(1+\frac{4M^2x^2}{Q^2}\right)^{1/2}\, ,  x_N=\frac{Q^2}{2(p^0q^0-p_zq_z)}.
\end{equation}
Here $F_{2,3}^{proton}$ and $F_{2,3}^{neutron}$ are the dimensionless structure functions for the free proton and the free neutron respectively.
These structure functions are calculated with target mass correction (TMC) \cite{schienbein} and 
CTEQ6.6 parton distribution functions (PDFs) at the Leading-Order (LO)~\cite{cteq}. Fermi motion and nucleon binding are implemented through the use of a nucleon spectral function. This is our base result. 
We also include pion and rho cloud contributions in $F^A_2$ following the model of Ref.~\cite{Marco:1995vb} and 
shadowing corrections in $F^A_2$ and $F^A_3$ \cite{Kulagin1}, which is our full calculation (Total).
Using them we have obtained the charged current differential scattering cross sections, the expression for which is given
 by Eq.~(\ref{cc_xsection}). For the neutral current the expression would remain the same with the changes given in Eq.~(\ref{nc_changes}).
 Using these cross sections, we evaluate the ratio
\begin{equation}\label{diff_ratio}
 R^-=\frac{\frac{d^2\sigma(\nu_\mu~N \rightarrow \nu_\mu~X)}{dxdy}~-~\frac{d^2\sigma(\bar\nu_\mu~N \rightarrow \bar\nu_\mu~X)}{dxdy}}
{\frac{d^2\sigma(\nu_\mu~N \rightarrow \mu^-~X)}{dxdy}~-~\frac{d^2\sigma(\bar\nu_\mu~N \rightarrow \mu^+~X)}{dxdy}}
\end{equation}

Results for the charged current induced (anti)neutrino-iron differential cross section have been discussed in detail in Ref.~\cite{Sajjad2},
where the comparisons have been made with the experimental data of NuTeV and CDHSW which are corrected for isoscalar iron target.
However, for the completeness we are showing here the results for the charged and neutral current induced (anti)neutrino-iron differential cross section
in Figs.\ref{d2sigma_cc_nu_nubar} and \ref{d2sigma_nc_nu_nubar} respectively. In the case of neutral current we have performed numerical calculations at LO.
We observe that medium effects are important in the study of differential scattering cross sections. For example, when the calculations are performed for charged current neutrino induced process
 at LO,
medium effects like Pauli blocking, Fermi motion and nucleon correlations which are taken into account for our base calculations, 
 results in the reduction of the differential cross section which changes by $3-4\%$ at low y and increases to $10-11\%$ at high values of y for the studied region of x
 from the free case. When pion and rho cloud contributions as well as shadowing effect are also taken into account there is a further change
 of about $8-9\%$ at low y and this change decreases with the increase in y for  x=0.2-0.3.
 This difference becomes smaller with the increase in the value of x, for example at x=0.5-0.6, it becomes less than $1\%$ for all values of y.
  Similarly, in the case of antineutrino induced charged current process the change in the base results from the free nucleon scattering process is around $10-12\%$ at x=0.2-0.3
  for studied region of y,
  which gets further modulated by  $12\%$ at low values of y and significantly increases with the increase in the value of y when 
pion and rho cloud contributions and shadowing effects are incorporated.
At higher values of x, the difference in the results obtained using the base calculation and the full calculation is negligibly small and the difference between the base results and the results 
for the free case is the same as in the case of neutrino. Furthermore, calculations performed at NLO leads to better results.
In the case of neutral current we have performed calculations at LO and observe that the difference between the base results and the results for the free case (not shown here) is $10-12\%$ at x=0.2-0.3 for neutrino
as well as antineutrino induced processes. When pion and rho cloud contributions and shadowing effects are taken into account, results from base change to $17-25\%$ for neutrino case and for antineutrino 
induced process the change is $30-40\%$  at x=0.2-0.3 for all values of y.
At higher x,  when Fermi motion, Pauli blocking and nucleon correlation are taken into account, the results change by $4-10\%$ from the free results at all values of y for
both neutrino as well as antineutrino induced processes. When pion and rho cloud contributions as well as shadowing effects are also considered, there is a 
further reduction of $2-3\%$ for neutrino induced process and $4-9\%$ for antineutrino case for all values of y. 
Therefore, we observe that in the study of charged current and neutral current differential scattering cross sections medium effects are important.

The PW ratio($R^-$), given by Eq.~(\ref{diff_ratio}),  is shown in Fig.\ref{rminus} using the numerical values from charged and neutral current differential scattering cross sections in iron, treating it to be
isoscalar as well as nonisoscalar nuclear targets, at different values of x for a fixed value of neutrino/antineutrino energy $E_{\nu(\bar \nu)}$=80 GeV. We have incorporated 
Fermi motion, Pauli blocking and nucleon correlation while calculating  $R^-$. 
We find that $R^-$ is almost independent of x and y for an isoscalar target, while for the  nonisoscalar target there is x as well as y dependence.
We find that the nonisoscalarity and medium effect increase with the increase in the value of x, for 
example in the mid region of x it is about  $14 - 15\%$ for low values of y and $3 - 4\%$ at higher values of y. 
Therefore, for nonisoscalar targets like iron, medium effects and excess of neutrons strongly affect $R^-$ and hence $sin^2\theta_W$. While for an isoscalar
target like carbon(not shown here) medium effects cancel out and extracted value of  $sin^2\theta_W$ from PW ratio is in complete agreement with the global value.
When the contributions of pion and rho clouds are also taken into account, 
 we find that these changes do not bring any difference in the results of PW ratio as obtained with our base calculations. Since shadowing is a low x phenomenon, we have not considered it in our present study of PW ratio~\cite{McFarland}.  
To see the effect of nonisoscalarity in the iron target, we have plotted $\delta R^-$
using Eq.~(\ref{deltaRminus}) at $E_{\nu(\bar \nu)}=80$ GeV in Fig.\ref{deltaR}.
 We find that the effect of nonisoscalarity is large at low y and high values of x which decreases with the increase in the value of y as well as this effect is smaller at low values of x.
  Hence there is a nonisoscalarity dependence in the determination of $sin^{2}\theta_W$. 
 Using the results of $R^-$ from Eq.~(\ref{diff_ratio}) and $\delta R^-$ from Eq.~(\ref{deltaRminus}),
 we have obtained  $sin^{2}\theta_W$ using Eq.~(\ref{correctioninPW}) and presented the results in Fig. \ref{sintheta}.
  The above calculated value of weak mixing angle 
  is now corrected for isoscalar target where we should also keep in mind that the medium effects are still present.
 We find that due to medium effects $sin^{2}\theta_W$ is different from the global fit,
 and this difference is $\approx 7\%$ when evaluated for low value of y at  x=0.2 and this decreases to $1\%$ at high values of y, while this change is $\approx 9\%$ when
  calculated for low y at x=0.6 and this reduces to $2\%$ at high values of y.
  Thus we observe that nonisoscalarity as well as medium effects like Pauli blocking, binding energy and Fermi motion are important while extracting  $sin^{2}\theta_W$.
  To see the effect of neutrino energy $E_\nu$ and $Q^2$ dependence on $sin^2\theta_W$, we have plotted in Fig.\ref{sintheta_EQ},  $sin^2\theta_W$
  as a function of x at various values of $E_\nu$ and $Q^2$.  We observe that at 
$E_\nu$=80 GeV and $Q^2=25$ GeV$^2$ it is almost close to the standard value at low values of x, and  the value of $sin^2\theta_W$ changes significantly with $E_{\nu(\bar \nu)}$, $Q^{2}$ and x. 
Therefore, while looking at the NuTeV results it is also important to know at what values of x, $E_\nu$ and $Q^2$ , 
the analysis was performed as a wide range of these variables could change considerably the value of $sin^{2}\theta_W$.
\section{Conclusions}\label{sec:concl}
To summarize our results, in this work we have studied the effects of the nuclear-medium on the
structure functions $F_{2}^{A}(x,Q^{2})$ and $F_{3}^{A}(x,Q^{2})$ in the iron
nucleus treating it to be isoscalar nuclear target and then we have made non-isoscalarity corrections. We have used many-body 
theory to describe the spectral function of a nucleon in an infinite nuclear-medium. The spectral function takes
into account the Fermi motion and the binding energy of the nucleons. The spectral function also includes nucleon correlations. 
Then to apply it to the case of finite nucleus we have used the local-density approximation.  Target mass
corrections have also been considered. We have used CTEQ~\cite{cteq} PDFs in the numerical evaluation.  We have taken into consideration the
effects of mesonic degrees of freedom, of shadowing and anti-shadowing for the calculation of $F_{2}^{A}$ and for $F_{3}^{A}$ shadowing and antishadowing effects only. 

With these structure functions, we have evaluated numerically the differential scattering cross sections for the charged and neutral current induced (anti)neutrino interactions on the iron target.
 These differential scattering cross sections are then used to study the nuclear medium effects and nonisoscalarity correction in the extraction 
of weak mixing angle sin$^2\theta_W$ using Paschos-Wolfenstein(PW) relation. 

Beginning with a model for non-symmetric nuclear matter, we have expanded the nuclear hadron tensor in an isoscalar part plus nuclear corrections which are, roughly speaking, 
proportional to the difference between the neutron and the proton density profiles. We have performed this expansion for all the nuclear structure functions which appear
 in the Paschos-Wolfenstein ratio for differential cross sections and obtained the pure isoscalar Paschos-Wolfenstein result plus nuclear corrections which depend on the phase space 
Bjorken variables.

We have also observed that the Paschos-Wolfenstein ratio for isoscalar nuclear structure functions is almost independent of $x$ and $y$, as it should be. 
However, when one considers the model for non-symmetric nuclei, such as $^{56}$Fe, one obtains a PW ratio which depends on the phase space variables. 
Furthermore, we have evaluated the first order \textit{nuclear correction} to the isoscalar PW ratio and found it to be important for a wide range of the phase space.
 Can this \textit{nuclear correction} explain the NuTeV anomaly is a question that should be answered if one could weight our results for $\delta R^{-}$ by the relative amount of 
events in every $x$ and $y$ bin, under the assumption of assigning all the correction to $\sin^2\theta_W$.

Another important point that may be taken into account is the fact that we have worked in the up and down quarks approximation. Therefore,
 we have neglected the contribution coming from heavier flavors. We leave this subject for future studies.
\begin{acknowledgments}
The authors thank M. J. Vicente Vacas, University of Valencia, Spain for many useful discussions and encouragement throughout this work. This research was supported by DGI
and FEDER funds, under contracts FIS2011-28853-C02-01, and the Spanish Consolider-Ingenio 2010
Program CPAN (CSD2007-00042), by Junta de Andaluc\'ia Grant No. FQM-225,
by Generalitat Valenciana contract PROMETEO/2009/0090 and by the EU
HadronPhysics2 project, Grant Agreement No. 227431. M. S. A. wishes to acknowledge the financial support under the grant SR/S2/HEP-0001/2008. H. H. acknowledges the Maulana Azad National Program. 
I. R. S. also wishes to acknowledge the financial support from MIUR grant PRIN-2009TWL3MX.
\end{acknowledgments}

\end{document}